\documentclass[a4paper,11pt]{article}

\usepackage{graphicx}
\usepackage{amsmath}
\usepackage{amssymb}
\usepackage{authblk}
\usepackage{fancyhdr}
\usepackage[titletoc]{appendix}
 
\usepackage{bm}
\usepackage{amsfonts}
\usepackage{epstopdf}
\usepackage{xcolor}
\usepackage{color}
\usepackage{wasysym}

\usepackage{dcolumn}
\usepackage{bm}
\usepackage{float}
\usepackage{tabularx}
\usepackage[a]{esvect}
\usepackage{subcaption}

\newcommand\clearrow{\global\let\rowmac\relax}
\clearrow

\title{\sffamily{Optimized setups for detection of Megatesla-level magnetic fields through Faraday rotation of XFEL beams}}

\author[1]{T. Wang}
\affil[1]{Department of Mechanical and Aerospace Engineering, University of California at San Diego, La Jolla, CA 92093, USA}

\author[2]{T. Toncian}
\affil[2]{Institute for Radiation Physics, Helmholtz-Zentrum Dresden-Rossendorf e.V., 01328 Dresden, Germany}

\author[3]{M. S. Wei}
\affil[3]{General Atomics, San Diego, California 92121, USA}

\author[1]{A. V. Arefiev}

\date{\today}

\begin{document}

\vskip -2.0cm
\maketitle
\vskip -3.0cm

\begin{abstract}
{A solid density target irradiated by a high-intensity laser pulse can become relativistically transparent, which then allows it to sustain an extremely strong laser-driven longitudinal electron current. The current generates a filament with a slowly-varying MT-level azimuthal magnetic field that has been shown to prompt efficient emission of multi-MeV photons in the form of a collimated beam required for multiple applications. This work examines the feasibility of using an x-ray beam from the European XFEL for the detection of the magnetic field via the Faraday rotation. Post-processed 3D particle-in-cell simulations show that, even though the relativistic transparency dramatically reduces the rotation in a uniform target, the detrimental effect can be successfully reversed by employing a structured target containing a channel to achieve a rotation angle of $10^{-4}$ rad. The channel must be relativistically transparent with an electron density that is lower than the near-solid density in the bulk. The detection setup has been optimized by varying the channel radius and the focusing of the laser pulse driving the magnetic field. We predict that the Faraday rotation can produce $10^3$ photons with polarization orthogonal to the polarization of the incoming 100 fs long probe beam with $5 \times 10^{12}$ x-ray photons. Based on the calculated rotation angle, the polarization purity must be much better than $10^{-8}$ in order to detect the signal above the noise level. 
}
\end{abstract}


\section{\sffamily{Introduction}}

Photon beams with energies in the multi-MeV energy range can have multiple biomedical~\cite{MV_x_rays_CT,MV_x_rays_therapy} and national security~\cite{NRF_3MeV,AI_homeland} applications and they can also open up new areas of fundamental research that heavily rely on photon collisions~\cite{Ribeye_pair_creation}. Motivated by these prospects, a significant effort has been directed towards identifying and exploring physics regimes that would enable efficient generation of gamma-rays~\cite{Chen_MeV_x_ray,MeV_nonlinear_Thomson,MeV_Geddes2015,jll_photon}. Some of the novel and efficient schemes rely on ultra-intense laser pulses that are expected to become available at the next generation of laser facilities that are currently under construction, such as the Extreme Light Infrastructure~\cite{ELI_NP}.  

One recently proposed regime that might be accessed at the already existing laser facilities involves extreme Megatesla-level magnetic fields that are volumetrically driven in an solid target irradiated by an intense laser pulse~\cite{Stark-PhysRevLett.116.185003,Oliver_pair}. Deflections of laser-accelerated electrons in such a strong field lead to emissions of energetic photons and result in an unprecedented efficiency. Three dimensional (3D) kinetic simulations predict that more than 3\% of the incoming laser energy can be converted into multi-MeV photons at laser intensities not exceeding $5 \times 10^{22}$ W$/$cm$^2$. 

The ultra-high laser intensity is the key to achieving this regime, since a strong electric field is required to make a significantly over-critical and otherwise opaque plasma relativistically transparent. Electron motion becomes ultra-relativistic in the strong laser field, so the electrons become effectively heavier and an opaque plasma becomes transparent to the laser pulse. The onset of relativistic transparency enables laser propagation through an over-critical plasma at solid densities and makes it possible to drive strong, but relatively slow-varying, azimuthal magnetic fields.

Successful detection of magnetic fields is the key to experimental validation of those regimes that critically depend on the presence of extreme magnetic fields in the laser-irradiated plasma, such as the regime of efficient gamma-ray generation. However, the combination of the unprecedented magnetic field strength and high plasma density rules out conventional optical and charged particle probing techniques as methods of measuring the magnetic fields of interest.

The goal of this work is to examine the feasibility of using an x-ray free electron laser (XFEL)~\cite{LCLS_2010,XFEL_review,EUXFEL,Japan_XFEL} as a magnetic field detection tool in the regimes of laser-plasma interaction similar to the one described above. In order to be quantitative, we perform our 3D particle-in-cell  (PIC) simulations and the subsequent post-processing analysis for the parameters that are projected for the high energy density (HED) station~\cite{hibef} at the European XFEL~\cite{EUXFEL}. We find that the polarization rotation of the x-ray beam is dramatically reduced in our regime of interest compared to what was predicted for a regime where the laser-plasma interaction is restricted to the surface of the target~\cite{Huang2017}. The cause of the reduction is the relativistically induced transparency that enables propagation of the laser pulse and its volumetric interaction with the plasma. We demonstrate that the detrimental effect of the relativistic transparency on the polarization rotation can be successfully mitigated by employing a structured target. The target structure leads to an electron density increase in the region with a strong magnetic field that is located at the periphery of the laser beam, while simultaneously reducing the characteristic electron relativistic $\gamma$-factor. We have further optimized this setup by varying the focal parameter of the laser pulse and the radius of the channel while keeping the laser pulse duration and its energy fixed. Our main conclusion is that an optimal setup can produce detectable polarization rotation of the probing x-ray beam even in the presence of the relativistic transparency. Therefore, XFEL beams present a viable option for detecting extreme laser driven magnetic fields (hundreds of kT in strength) that are embedded in a dense optically opaque plasma and driven by a 300 TW level laser.


\section{\sffamily{Assessment of the established magnetic field detection techniques}}

Successful detection of magnetic fields is the key to experimental validation of those regimes that critically depend on the presence of such fields inside a plasma. Our regime of interest has two distinct features. The magnetic field is at least an order of magnitude stronger than the fields that have been previously detected. In addition to that, the magnetic field is embedded in a plasma whose density is almost two orders of magnitude higher than the critical density for an optical laser, which means that this plasma is opaque. In what follows, we assess the feasibility of using common detection techniques for this specific regime and show that a different approach is required. 

There are two approaches that have been successfully applied in the past to detect transient magnetic fields generated during laser-plasma interactions. One approach is to utilize the deflection of charged particles (protons~\cite{law2016_kT_proton,santos2015_800T_proton_FR} or electrons~\cite{Schumaker_PRL_2013}) caused by the fields. The other approach is to utilize a magneto-optical phenomenon known as the Faraday effect~\cite{stamper1975_seminal_faraday,willi1998,kaluzaPRL2010,walton2013_FR_experiments_abel}. This effect manifests itself as a rotation of the polarization plane in the transmitted light, with the rotation proportional to the strength of the magnetic field. Out of these two approaches, only the one that involves energetic particles can be applied to detect the magnetic field in our regime. The optical probing is automatically ruled out since the field is surrounded by an optically opaque material. 

Probing plasma electric and magnetic fields using laser-driven proton beams is one of the more mature and, as a result, frequently used techniques. Laser-accelerated proton beams have unique properties, such as the ps-scale duration and high laminarity, that make them ideal for measurements that require simultaneous temporal (ps scale) and spatial (tens of micrometers) resolution. As MeV-level protons penetrate an overcritical plasma, they are deflected by electric and magnetic fields while their trajectories remain relatively unaffected by binary collisions. The plasma field topology is then reconstructed from the resulting proton deflection image~\cite{cecchetti2009_45T_proton}.  A key advantage of using protons is that they enable probing of optically opaque (overcritical) plasmas. This technique has been applied to probe magnetic fields that develop in subcritical (optically transparent) plasmas. It has been used to complement optical Faraday rotation measurements, which enabled successful detection of magnetic fields with a strength in the kT range~\cite{law2016_kT_proton,santos2015_800T_proton_FR}. 

The key parameters in our regime of interest differ in two significant ways from those encountered in experiments where the proton probing was used successfully: the strength of the magnetic field is significantly higher, while the transverse size of the volume occupied by the magnetic field is much smaller. Detecting a MT-level magnetic field would require protons with energies in the range of 100 MeV, since this magnetic field is almost two orders of magnitude stronger than what is being currently detected. Laminar laser-driven proton beams with these energies are not yet available. The transverse size of the magnetic field that needs to be detected is roughly 10 $\mu$m~\cite{Stark-PhysRevLett.116.185003,Oliver_pair}. It is comparable to the virtual source size of the available proton beams. Therefore, the detection technique involving such beams lacks the required resolution. 

High energy relativistic electrons might be an attractive option for probing the regime of interest due to their extended propagation length in solid density plasmas, but the energies required to probe MT-level magnetic fields appear to be beyond the capabilities of what is typically achievable using laser-plasma accelerators. We arrive to this conclusion based on the fact that, according to our kinetic simulations~\cite{Stark-PhysRevLett.116.185003,Oliver_pair}, the laser-driven magnetic field that we intend to measure easily confines electrons with energies approaching 1 GeV. Therefore, electron beams with energies much higher than 1 GeV would be needed to image the magnetic field structure. 

Evidently, none of the available techniques are well-suited for detecting MT-level magnetic fields in solid-density plasmas. It is therefore paramount to investigate and develop alternative options and such an investigation is the objective of this paper. 


\section{\sffamily{Preliminary assessment of x-ray Faraday rotation\\as a detection technique}} \label{Sec-3}

X-ray free electron lasers, such as LCLS at SLAC~\cite{LCLS_2010}, SACLA at Riken Harima~\cite{Japan_XFEL}, and now the European XFEL~\cite{EUXFEL}, offer an intriguing alternative of using x-ray photons to detect strong magnetic fields inside a dense plasma. For example, an x-ray beam of 6 keV photons can easily penetrate a plasma whose electron density is 100 times  higher than the critical density for an optical laser pulse with a 1 $\mu$m wavelength. Conceptually, this approach would be similar to that used for optical probing of magnetic fields in plasmas~\cite{stamper1975_seminal_faraday,willi1998,kaluzaPRL2010,walton2013_FR_experiments_abel}. The polarization plane of a linearly polarized x-ray beam would rotate as it passes through a dense plasma with a magnetic field. The rotation angle scales as $\Delta \phi \propto B_{\parallel}  n_e \Delta l$, where $n_e$ is the electron density, $\Delta l$ is the thickness of the plasma that contains the magnetic field, and $B_{\parallel}$ is the strength of the magnetic field component directed along the path traveled by the beam. The rotation angle would thus contain the information about the strength of the magnetic field that one can then try to recover. 


The ability of an x-ray beam to easily probe a magnetic field inside a very dense plasma is a direct consequence of its frequency $\omega_*$ being much higher than the frequency $\omega$ of an optical laser beam. The increased transparency however comes at a cost. The rotation angle scales as $\Delta \phi \propto 1/ \omega^2_*$, so the rotation angle drops as the frequency increases. Clearly, the key issue for this approach is the strength of the detectable signal. It has to be evaluated for a specific experimental setup to determine the feasibility of the approach. 

One such study has been recently performed to assess the detection of kT-level magnetic fields~\cite{Huang2017} driven by the HiBEF \cite{hibef} High Intensity optical laser at the European XFEL. In the considered setup, numerous magnetic field filaments are driven inside a solid density plasma by streaming hot electrons, launched into the target by the laser pulse. The magnetic field is primarily azimuthal in these filaments whose typical transverse size is 0.4 $\mu$m. It was calculated that the corresponding Faraday rotation for a 6.457 keV x-ray beam of the European XFEL \cite{EUXFEL} is several hundred $\mu$rad. This rotation is accumulated as the x-ray beam traverses multiple filaments in the plasma. The authors of the study have deduced that the calculated Faraday rotation should be detectable via x-ray polarimetry~\cite{Siddons1990_x_ray_Faraday} using Si channel-cut crystals as polarizers. The x-ray signal would have to be integrated over tens of shots in order to obtain a good signal to noise ratio.

It is worth pointing out that the polarization of the probe beam changes not only if it propagates parallel to the magnetic field lines (Faraday rotation). The polarization changes even if the propagation direction is normal to the field lines, which is often referred to as the Cotton-Mouton effect. We find that for the 6.457 keV x-ray probe beam and a 0.4 MT magnetic field the strongest rotation occurs when the x-ray beam propagates parallel to the magnetic field lines, which is the Faraday rotation (see Appendix~\ref{App_A}). In the case of a cold plasma, the Cotton-Mouton effect is smaller by a factor of $\omega_{ce} / \omega_* \approx 7 \times 10^{-3}$, where $\omega_{ce}$ is the electron cyclotron frequency for the considered magnetic field. This is the reason why we focus on the Faraday rotation in the remainder of the paper.

\begin{figure} [H]
   \subcaptionbox*{}
      [0.32\linewidth]{\includegraphics[scale=0.32,trim={0.5cm 1.7cm 2cm 0.5cm},clip]{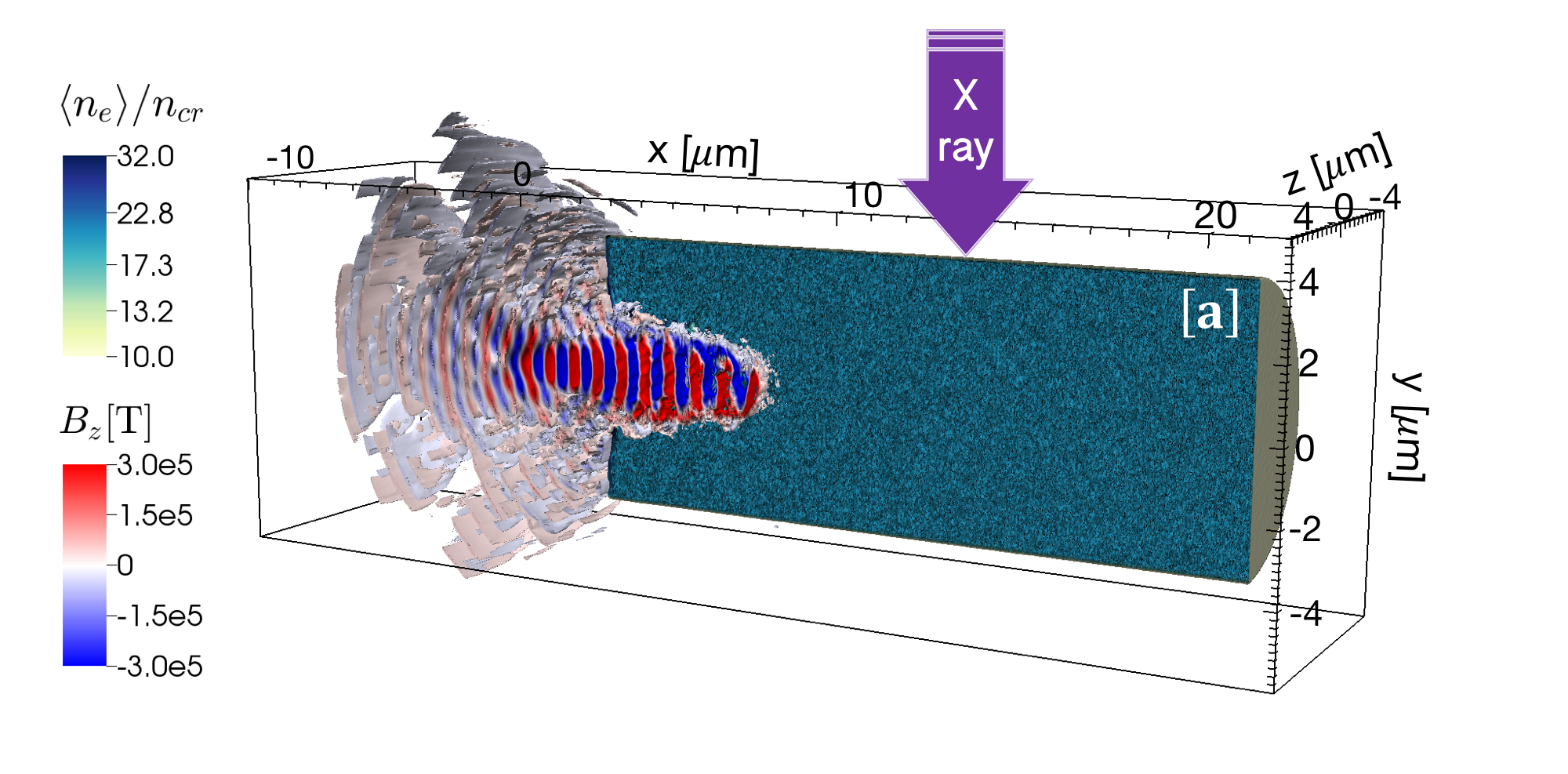}}
      \hspace*{45mm}
   \subcaptionbox*{}
      [0.3\linewidth]{\includegraphics[scale=0.23]{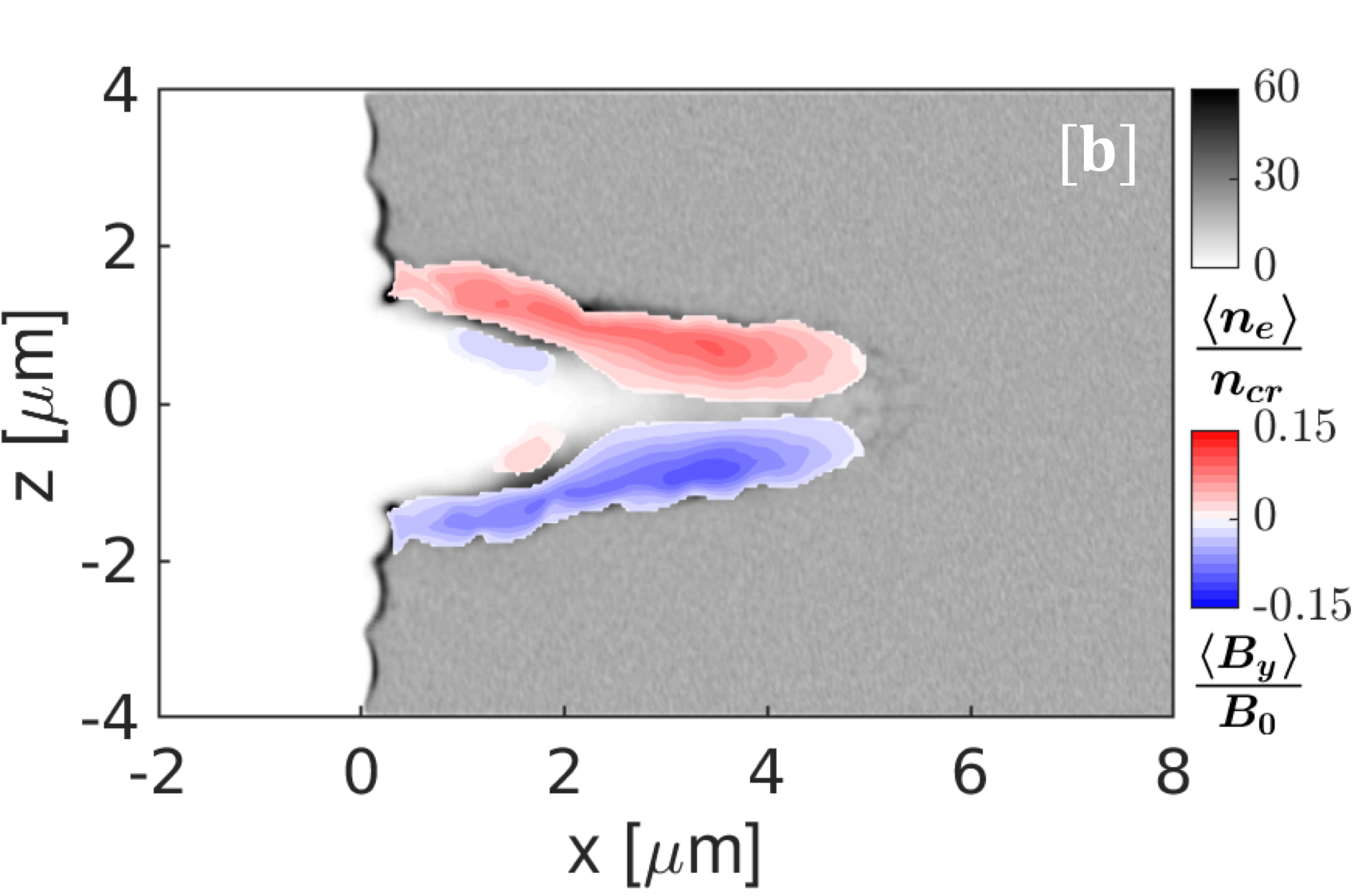}}
      \caption{\label{FIG-sim_setup_uniform}  Generation of a strong quasi-static magnetic field in an initially uniform target with $n_e = 20 n_{cr}$ using relativistically induced transparency. The laser pulse penetrates the target (a) and drives an azimuthal slowly-evolving magnetic field (b). The left panel (a) shows the instantaneous magnetic field that includes the field of the laser and the field of the plasma, whereas the right panel (b) shows the time-averaged magnetic field only.}
\end{figure}

This preliminary discussion indicates that, in principle, the Faraday rotation of x-ray beams offers a path to detecting magnetic fields in optically opaque plasmas. 
An XFEL beam posses a combination of spatial and temporal resolutions needed to detect $\mu$m-size magnetic structures that exist for less than 100 fs is required. 



\section{\sffamily{Preliminary assessment of the role played\\by the relativistic transparency}}

Our regime of interest has an important feature that distinguishes it from  the regime considered in the published feasibility study~\cite{Huang2017} and described in Section~\ref{Sec-3}. This feature is the relativistically induced transparency that enables an ultra-intense laser pulse to propagate through a dense plasma and drive a strong longitudinal electron current~\cite{Stark-PhysRevLett.116.185003, Oliver_pair, ji2014_Megatesla, qiao2017_Megatesla}. The magnetic field is generated and supported by this current in a region with a very energetic bulk electron population, as opposed to being generated in an effectively cold plasma~\cite{Huang2017}. In this Section, we examine how the presence of the heated bulk electrons affects the Faraday rotation of an x-ray probe beam.

Figure~\ref{FIG-sim_setup_uniform}a shows a generic setup that can be used to generate a strong magnetic field using relativistically induced transparency. The key is to correctly choose the target material for a given peak electric field amplitude $E_0$ of the irradiating laser pulse. It is convenient to use the so-called normalized laser amplitude, $a_0 = |e| E_0 / m_e c \omega$, to make this assessment. Here $m_e$ and $e$ are the electron mass and charge, while $\omega$ is the frequency of the laser pulse. The plasma is transparent if the electron density $n_e$ satisfies the condition $n_e \ll \gamma_{av} n_{cr}$, where $\gamma_{av}$ is the characteristic (average) relativistic factor of the bulk electron population and $n_{cr}$ is the classical critical density defined as $n_{cr} \equiv \omega^2 m_e / 4 \pi e^2$. Typically, the $\gamma$-factor resulting from bulk electron heating by the laser can be estimated as $\gamma_{av} \approx a_0$. Then the condition that must be satisfied by the target electron density in order for the target to become relativistically transparent is 
\begin{equation} \label{RT}
n_e \ll a_0 n_{cr} = \frac{a_0 m_e \omega^2}{ 4 \pi e^2}.
\end{equation} 
It is helpful to view the increased transparency as a result of an effective mass increase, with the electron mass ``increasing'' to $\gamma_{av} m_e \approx a_0 m_e$ due to the relativistic motion.

The fact that the plasma becomes transparent to the irradiating laser beam means that it would also become more transparent to the probing x-ray beam. The heating is therefore likely to significantly reduce the Faraday effect and, as a result, the polarization rotation compared to the case where the plasma electrons are assumed to be cold. In order to provide a preliminary assessment of this effect, we consider a simple case of a uniform plasma with cold electrons. The magnetic field $B_{\parallel}$ is also uniform and it is probed by a linearly polarized x-ray beam propagating along the magnetic field lines. In this case, the polarization rotates by
\begin{equation}
\Delta \phi_{cold} = \frac{1}{2} \frac{n_e}{n_*} \frac{|e| B_{\parallel}}{m_e c} \frac{\Delta l}{c},
\label{FR_cold} 
\end{equation}
where $\Delta l$ is the distance traveled by the probe beam and 
\begin{equation} \label{n_*-def}
n_* \equiv \frac{\omega_*^2 m_e}{4 \pi e^2}    
\end{equation} 
is the critical density for the considered x-ray beam with frequency $\omega_*$. We can now estimate the polarization rotation in a plasma with relativistically hot electrons, $\Delta \phi_{hot}$. We account for the relativistic motion by changing the electron mass to $\gamma_{av} m_e$ in Eq.~(\ref{FR_cold}), which represents the effective mass increase mentioned earlier, and find that
\begin{equation}
\Delta \phi_{hot} \approx \frac{1}{2 \gamma_{av}^2} \frac{n_e}{n_*} \frac{|e| B_{\parallel}}{m_e c} \frac{\Delta l}{c} \approx \Delta \phi_{cold} \left/ \gamma_{av}^2 \right.,
\label{FR_hot} 
\end{equation}
where we also took into account that the classical critical density $n_*$ increases by a factor of $\gamma_{av}$. This estimate roughly matches the expression that has been rigorously derived for a relativistic electron population with a Maxwell-J\"uttner momentum distribution~\cite{FR_rel2,FR_rel3,FR_rel4,FR_Trubnikov_phd,FR_melrose_1997}. Appendix~\ref{App_A} provides an exact expression for $\Delta \phi_{hot}$ together with particle-in-cell simulation results performed for a wide range of $\gamma_{av}$ values. 

The obtained estimate indicates that we should expect for the polarization rotation to be dramatically reduced in our regime of interest that involves relativistic transparency. Indeed, the rotation decreases by two orders of magnitude even for a driving laser pulse with $a_0 \approx 10$, since $\gamma_{av} \approx a_0$. Some of our previous studies used an even higher laser amplitude, with $a_0 > 100$, to achieve a significant increase in the gamma-ray yield~\cite{Stark-PhysRevLett.116.185003, Oliver_pair}. Based on the simple estimate given by Eq.~(\ref{FR_hot}), the rotation in these regimes should drop by four orders of magnitude. 

To conclude this section, we use typical parameters observed in PIC simulations of the relativistically transparent regime~\cite{Stark-PhysRevLett.116.185003, Oliver_pair} to estimate the polarization rotation $\Delta \phi_{hot}$ for a 6.457 keV XFEL beam~\cite{Huang2017,MARX2011915}. We assume that the magnetic filament that causes the polarization rotation is 5 $\mu$m thick. The magnetic field is 400 kT and the electron density is 20 $n_{cr}$, where $n_{cr}$ corresponds to a driving laser pulse with a wavelength of 800 nm. The wavelength of the x-ray beam is 0.192 nm, so that $n_{cr} / n_* = (0.192 / 800)^2 \approx 5.8 \times 10^{-8}$. It then follows from Eq.~(\ref{FR_hot}) that $\Delta \phi_{hot} \approx 5.1 \times 10^{-8}$ rad. In contrast to that, the same filament would rotate the polarization by $\Delta \phi_{cold} \approx 5.5 \times 10^{-4}$ rad if the electrons were cold. 

Our estimate for the rotation in a relativistically transparent plasma provides a useful baseline, but it should be viewed as ``the worst case scenario'' because it ignores transverse variations of $\gamma_{av}$. In the case of a relativistically transparent plasma filament embedded in a cold over-critical plasma, the average relativistic factor $\gamma_{av}$ changes dramatically across the filament. It falls from its maximum value in the center where the laser pulse has the strongest field to $\gamma_{av} \approx 1$ at the periphery of the laser beam. Moreover, the azimuthal magnetic field that increases towards the periphery of the filament, which should further affect the rotation angle. In the next section, we explore the impact that these variations have on the rotation angle and show that they can be used to increase it. 

\begin{table}
\centering
\small
\vspace*{-2em}
\caption{Parameters of 3D PIC simulations}
\label{table_PIC}
\begin{tabular}{ |l|l| }
  \hline
  \multicolumn{2}{|l|}{\textbf{Laser pulse parameters:} }\\
  Peak intensity & $8 \times 10^{22}$ W/cm$^2$ \\
  Wavelength & 0.8 $\mu$m \\
  Energy injected & 9.04 J \\
  Pulse duration (FHWM for intensity) & 30fs\\
  Focal spot size (FWHM for intensity) & 0.59 $\mu$m \\
  Location of the focal plane & $x$ = 0 $\mu$m\\
  \multicolumn{2}{|c|}{}\\
  \multicolumn{2}{|l|}{\textbf{General parameters: }}\\
  Length and radius of cylindrical target & 20.0 $\mu$m and  4.0 $\mu$m\\
  Spatial resolution & $50/\mu m\times 20/\mu m\times 20/\mu m$ \\
  Number of macro-particles/cell (for each species) & 10 \\
  \multicolumn{2}{|c|}{}\\
  \multicolumn{2}{|l|}{\textbf{Uniform target: }}\\
  Ion species & $H^{+1}$ \\
  Electron and ion densities & $n_e = n_{H^{+1}} = 20n_{cr}$ \\
  \multicolumn{2}{|c|}{}\\
  \multicolumn{2}{|l|}{\textbf{Target with a cylindrical channel: }}\\
  Channel radius  & 0.8 $\mu$m \\
  Channel ion species & $H^{+1}$ \\
  Electron and ion densities in the channel & $n_e = n_{H^{+1}} = 20n_{cr}$ \\
  Bulk ion species & $C^{+6}$ \\
  Electron and ion densities in the bulk & $n_e = 6n_{C^{+6}} = 100n_{cr}$ \\
  \hline
\end{tabular}
\end{table}


\section{\sffamily{Impact of relativistic transparency\\on x-ray Faraday rotation in a uniform target}} \label{Sec-uniform}

Electron heating is inherently nonuniform in the regime where relativistic transparency plays a central role and such a nonuniformity can significantly impact the polarization rotation of the x-ray beam. In order to examine this effect, we have carried out three-dimensional (3D) particle-in-cell (PIC) simulations that provide detailed information about spatial profiles of the characteristic $\gamma$-factor, $n_e$, and the magnetic field that we are interested in probing with the x-ray beam. Detailed simulation parameters are given in Table \ref{table_PIC}.

Figure~\ref{FIG-sim_setup_uniform} shows snapshots from a simulation where a high-intensity laser pulse propagates through an initially uniform over-critical plasma with $n_e = 20 n_{cr}$  while generating a strong azimuthal magnetic field. The peak intensity of the incoming laser pulse in the absence of the target is $8 \times 10^{22}$ W/cm$^2$. We choose this value because it is an upper limit for the high-intensity optical laser at the HiBEF HED station based on its specifications. The laser wavelength is 800 nm, so the normalized laser amplitude is $a_0 \approx 200$. In agreement with the condition given by Eq.~(\ref{RT}), the target becomes transparent when irradiated by this laser pulse. 

In this simulation, the instantaneous magnetic field (see Fig.~\ref{FIG-sim_setup_uniform}a) reaches a maximum value of 2 MT, which includes the field of the incoming and reflected pulses and the field generated by the plasma. The maximum amplitude without the plasma would have been $B_0 \equiv 2.6$ MT in the focal plane located at $x = 0$ $\mu$m. The laser pulse drives a magnetic field in the plasma that is evolving on a time-scale comparable to the laser pulse duration rather than rapidly oscillating with the laser frequency. The structure of this field is shown in Fig.~\ref{FIG-sim_setup_uniform}b. It is calculated by time-averaging the total magnetic field over four laser periods, with the averaging indicated by the angular brackets. The amplitude of this slowly-evolving field is $0.15 B_0$ or 0.4 MT, which confirms that the relativistically induced transparency indeed enables volumetric generation of a strong azimuthal magnetic field.

\begin{figure} [H]
   \begin{center}
      \includegraphics[scale=0.46]{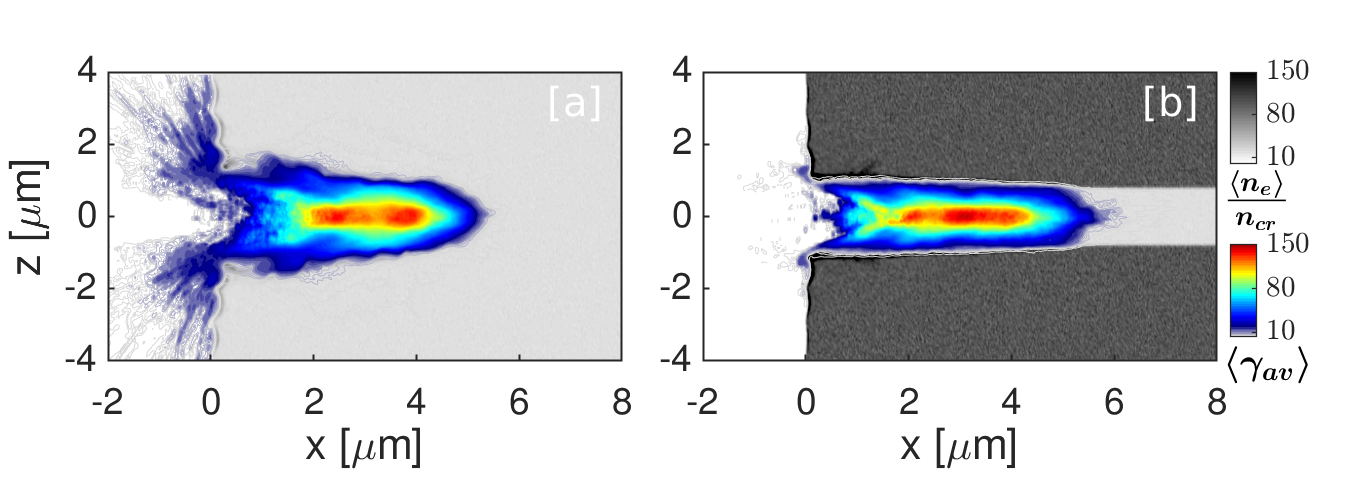} 
      \caption{\label{FIG-FR_gamma_both} Time-averaged relativistic $\gamma$-factor from a 3D PIC simulation for (a) an initially uniform  target and (b) a target with a pre-filled cylindrical channel. The gray-scale shows the time-averaged electron density in the target cross-section. Here $\langle \gamma_{av} \rangle $ is a cell-averaged value of $\gamma$. }
   \end{center}
\end{figure}

Figure~\ref{FIG-FR_gamma_both}a shows a snapshot of $\langle \gamma_{av} \rangle $, where the cell-averaged $\gamma$-factor that we denote as $\gamma_{av}$ has been averaged over four preceding laser periods. In what follows, the time-averaged values are indicated by angular brackets. The snapshot in Fig.~\ref{FIG-FR_gamma_both} is indicative of the electron heating induced by the propagating laser pulse. A comparison of the $\langle \gamma_{av} \rangle $ profile in Fig.~\ref{FIG-FR_gamma_both}a with the profile of $\langle B_y \rangle$ in Fig.~\ref{FIG-sim_setup_uniform}b reveals that the characteristic $\gamma$-factor changes significantly across the region with the strong azimuthal magnetic field. 

In order to quantify the impact of the electron heating, we have post-processed our simulation results in two different ways: 1) neglecting the electron heating, thus assuming that the electrons are cold; and 2) accounting for the electron heating by adjusting the polarization rotation using $\langle \gamma_{av} \rangle$. The Faraday rotation is calculated for a 6.457 keV XFEL beam~\cite{Huang2017,MARX2011915}. In all our post-processing calculations, the x-ray beam propagates along the $y$-axis, as shown in Fig.~\ref{FIG-sim_setup_uniform}a. The electric field of the incoming x-ray beam has only a $z$-component. The x-ray polarization rotation is found by evaluating the following integral along $y$ for all values of $(x,z)$ on our grid:
\begin{equation} \label{brot_correct} 
\Delta \phi (x,z) = \int_{y_{\min}}^{y_{\max}} \frac{\alpha (x,z,y)}{2} \frac{n_e (x,z,y)}{n_*} \frac{|e| B_y (x,z,y)}{m_e c} \frac{d y}{c},
\end{equation}
where $n_*$ is the classical critical density for the x-ray beam defined by Eq.~(\ref{n_*-def}). It is convenient to express it in terms of the critical density $n_{cr}$ for the irradiating laser pulse: $n_* \approx 1.74 \times 10^7 n_{cr}$. The function $\alpha(x, y, z)$ represents the change of the optical properties due to electron heating. If the electrons are non-relativistic, then $\alpha = 1$. Appendix~\ref{App_A} provides a compact analytical expression for $\alpha$ as a function of $\gamma_{av}$ that was confirmed using particle-in-cell simulations.

\begin{figure} [H]
   \begin{center}
      \includegraphics[scale=0.46] {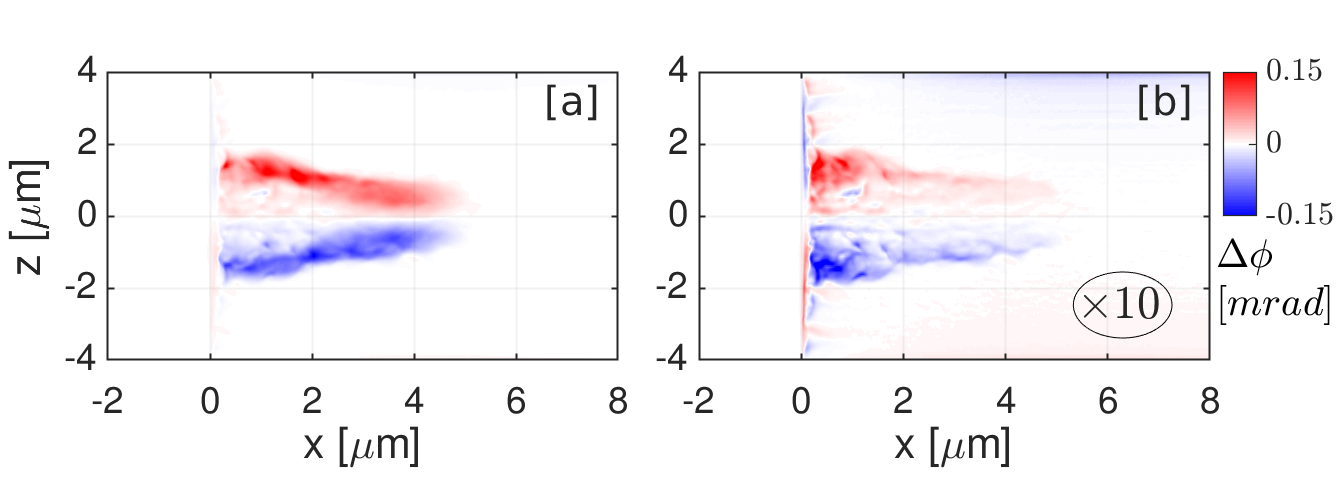} 
      \caption{\label{FR_angles_uniform} Faraday rotation of the x-ray probe beam that traverses the magnetic field filament shown in Fig.~\ref{FIG-sim_setup_uniform}. In panel (a), the rotation is calculated by assuming that the electrons are cold. In panel (b), the rotation is calculated by accounting for the electron heating to relativistic energies.
      The signal, i.e. the rotation angle, in panel (b) has been multiplied by a factor of ten in order to make it visible.}
\end{center}
\end{figure}

Figure~\ref{FR_angles_uniform}a shows the rotation angle calculated under the assumption that the plasma electrons are non-relativistic, i.e. by setting $\alpha(x, y, z) = 1$ in Eq.~(\ref{brot_correct}). The maximum polarization rotation exceeds 0.1 mrad. According to the previously published research~\cite{Huang2017}, this level of rotation might in principle be experimentally detectable in a single-shot experiment. The ``hollow'' spatial distribution of the signal is typical for the azimuthal field that we are probing. The signal peaks away from $z = 0$ because only $B_y$ contributes to the polarization rotation in our setup and this component increases away from $z = 0$.

\begin{figure}[H]
   \subcaptionbox*{}
      [0.32\linewidth]{\includegraphics[scale=0.32,trim={0.5cm 1cm 2cm 1.5cm},clip]{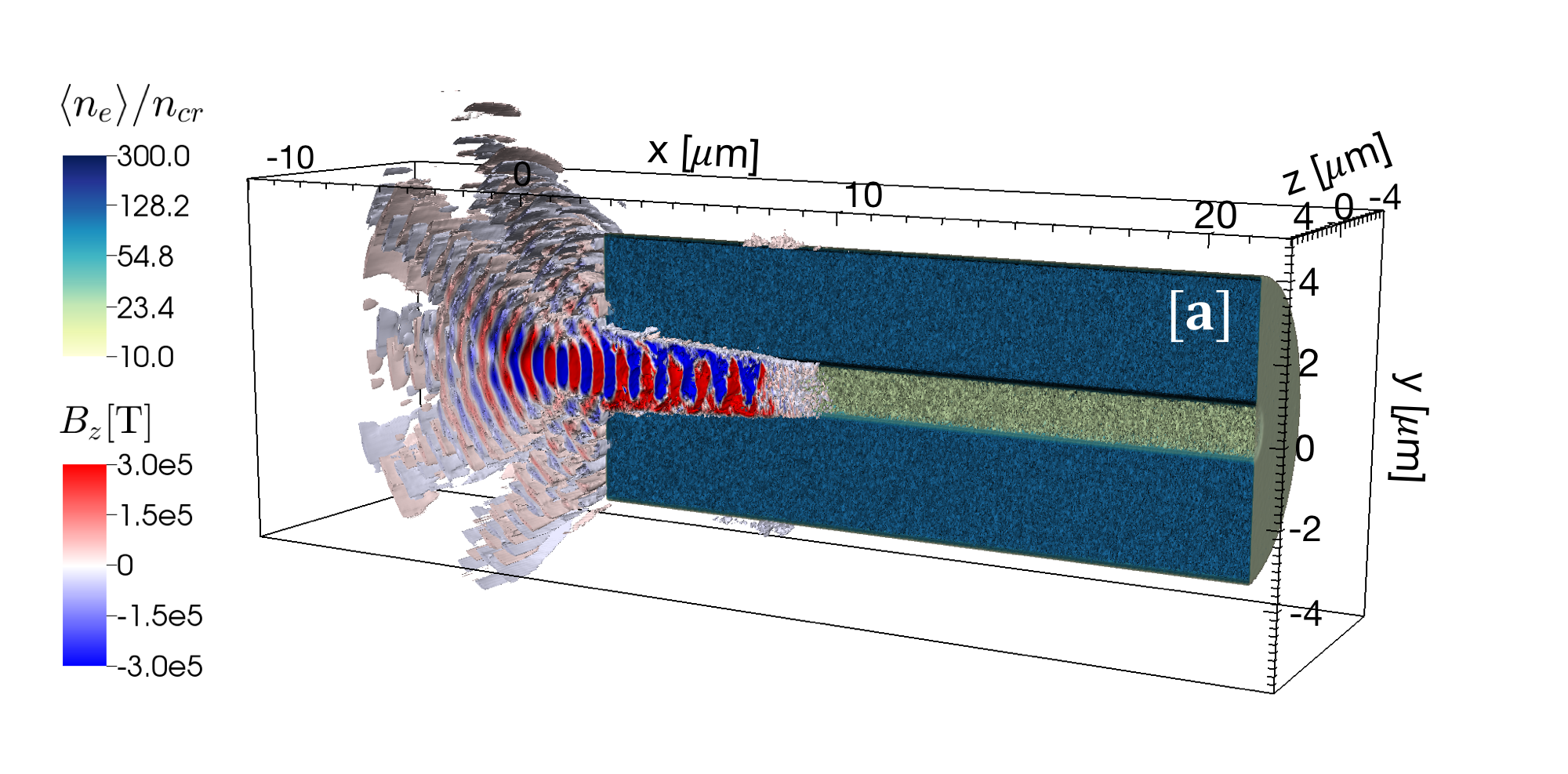}}
      \hspace*{45mm}
   \subcaptionbox*{}
      [0.3\linewidth]{\includegraphics[scale=0.23]{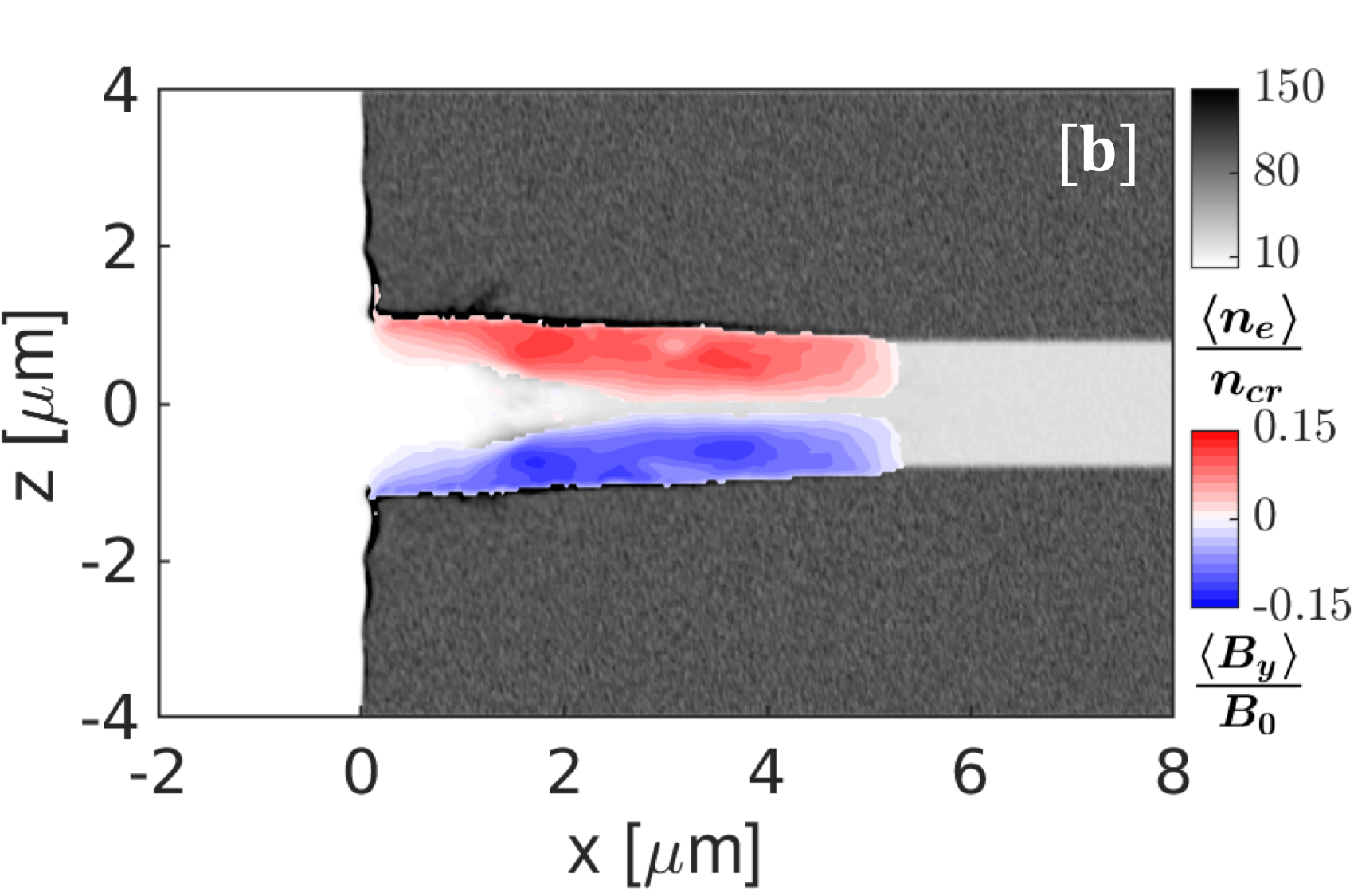}}
    \caption{\label{FIG-sim_setup_channel} Snapshots from a 3D PIC simulation of a structured target with a cylindrical channel irradiated by a high-intensity laser pulse. Initially, the electron density in the channel is $20 n_{cr}$, whereas the electron density in the bulk of the target is $100 n_{cr}$. The instantaneous magnetic field is shown in (a), whereas the time-averaged magnetic field is shown in (b). In both cases, the electron density is time-averaged.}
\end{figure}

Figure~\ref{FR_angles_uniform}b gives the rotation calculated by taking into account the heating of the electrons by the laser. As expected, the rotation angle is significantly reduced throughout the domain where the regions with high $\langle \gamma_{av} \rangle$ (see Fig.~\ref{FIG-FR_gamma_both}a) suppress the polarization rotation. The maximum rotation angle is only about 10 $\mu$rad and it is observed close to the surface of the target. Such a low signal is beyond a single-shot detection capability. In order to be detected, it would have to be reconstructed by integration over a large number of shots.


\begin{figure}[H]
	\centering
    \includegraphics[scale=0.46]{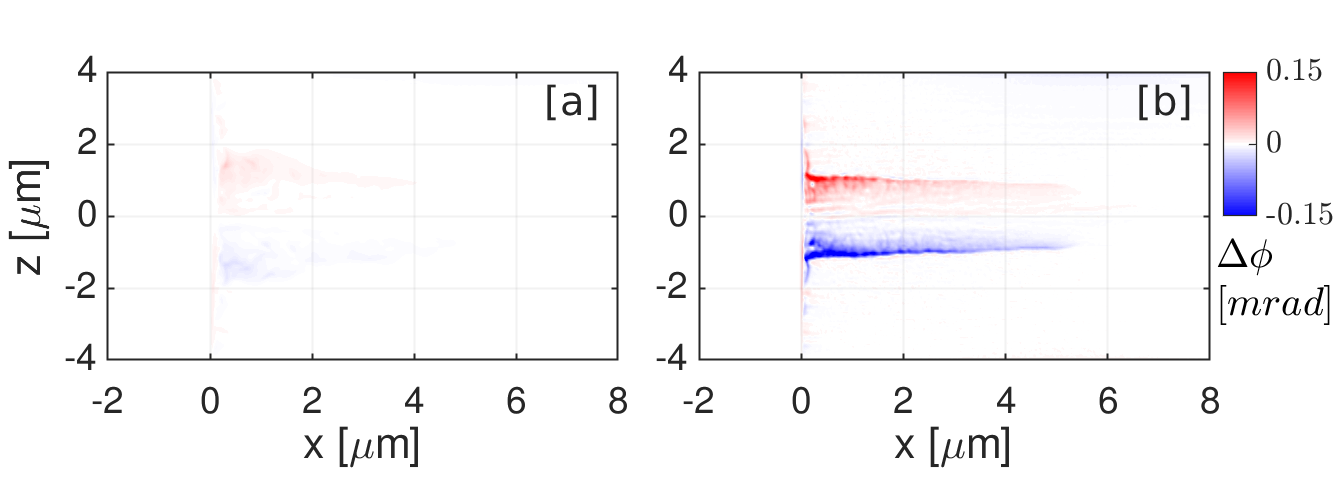}
    \caption{Faraday rotation of the x-ray probe beam caused by a strong magnetic field in a uniform (a) and in a structured target (b). Electron heating is taken into account in both cases.}
    \label{FR_w_wo_channel}
\end{figure}

\section{Increased Faraday rotation in a structured target} \label{Sec-5}

Our analysis in Sec.~\ref{Sec-uniform} has confirmed the initial assessment that the magnetic field detection is much more challenging in the presence of the relativistically induced transparency that dramatically reduces the polarization rotation. However, the rotation can be enhanced by employing structured targets. Our previous research has already showed that structured targets are beneficial for gamma-ray generation~\cite{Stark-PhysRevLett.116.185003,Oliver_pair}. In what follows, we demonstrate that they can also be used to make the magnetic field detection less challenging, as compared to the case of a uniform target.

\begin{figure}[H]
	\centering
    \includegraphics[scale=0.4,trim={1cm 2cm 0cm 2cm},clip]{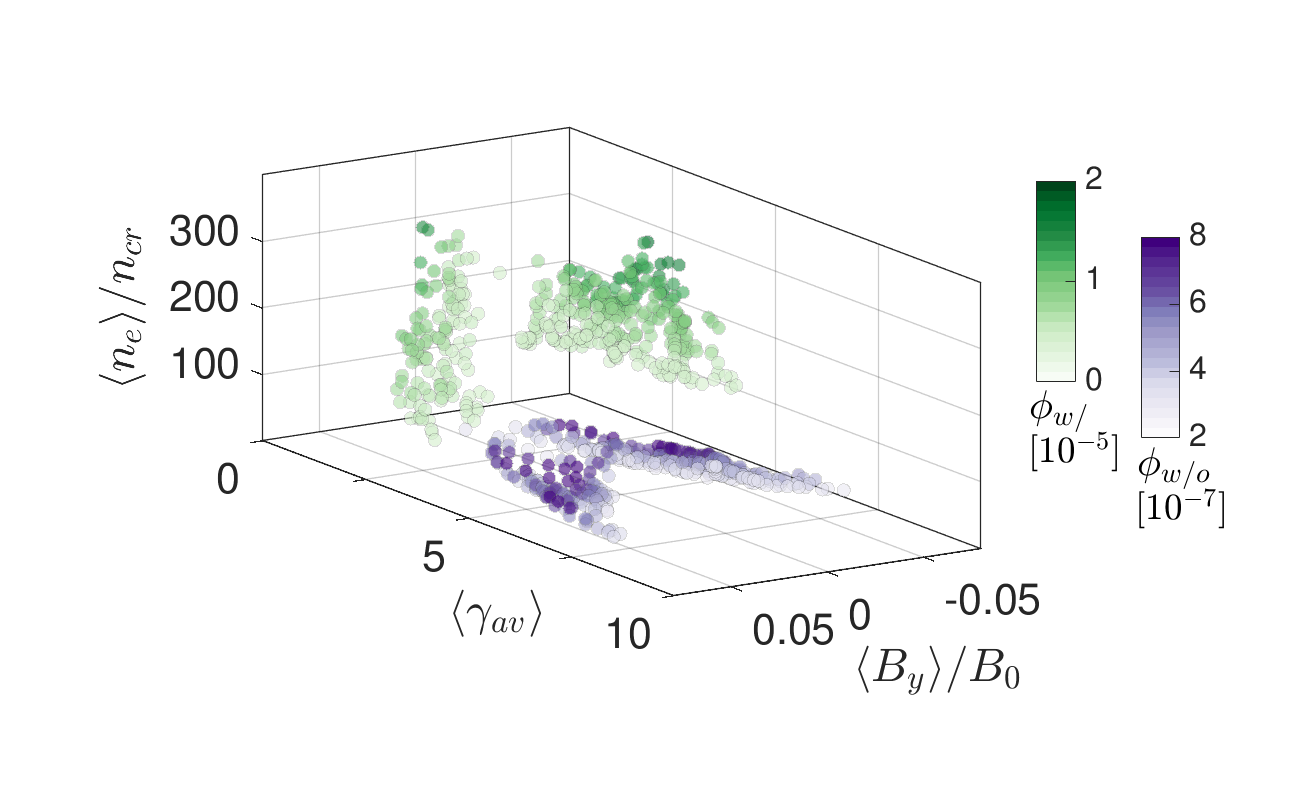}
    \caption{Faraday rotation induced by a thin plasma slab in a target with (green scale) and without (purple scale) a channel as a function of $\langle \gamma_{av} \rangle$, $\langle n_e \rangle$, and $\langle B_y \rangle$.}
    \label{FR_dependence}
\end{figure}

A structured target that we use in our simulations consists of a bulk material with electron density of $100n_{cr}$ and a narrow channel with electron density of $20 n_{cr}$. We assume that the material of the bulk and the material of the channel become fully ionized prior to the arrival of the intense laser pulse. The material of the bulk is plastic, represented by carbon ions. The material of the channel is hydrogen. Detailed simulation parameters are given in Table \ref{table_PIC}. Figure~\ref{FIG-sim_setup_channel} shows a snapshot from our 3D PIC simulation with the structured target. The channel becomes much more transparent than the bulk when irradiated by the intense laser pulse and this is what leads to the observed guiding of the laser pulse.

Figure~\ref{FR_w_wo_channel}b shows the polarization rotation of the x-ray probe beam after it has traversed the structured target. The post-processing is performed by taking into account electron heating by the propagating laser pulse. For comparison, Fig.~\ref{FR_w_wo_channel}a shows the polarization rotation from the simulation with a uniform target. The rotation is increased by an order of magnitude in the case of the structured target, which clearly shows the advantage of using such a target. 

There are three factors that can potentially contribute to the increased rotation: 1) increased magnetic field, 2) reduced characteristic electron energy, i.e. $\langle \gamma_{av} \rangle$, in the region with a strong magnetic field, and 3) increased electron density in the region with a strong magnetic field. There is no significant increase in the maximum value of $\langle B_y \rangle$, so the target structure enhances the signal without significantly altering the field that we are trying to detect. This enhancement is caused by relative changes in the spatial profiles of $\langle \gamma_{av} \rangle$ and $\langle n_e \rangle$ with respect to $\langle B_y \rangle$. Indeed, there is no global reduction of $\langle \gamma_{av} \rangle$ (see Fig.~\ref{FIG-FR_gamma_both}) and the initial electron density in the channel is equal to the initial electron density in the uniform target.

\begin{figure*}[!ht]
	\centering
    \includegraphics[scale=0.5]{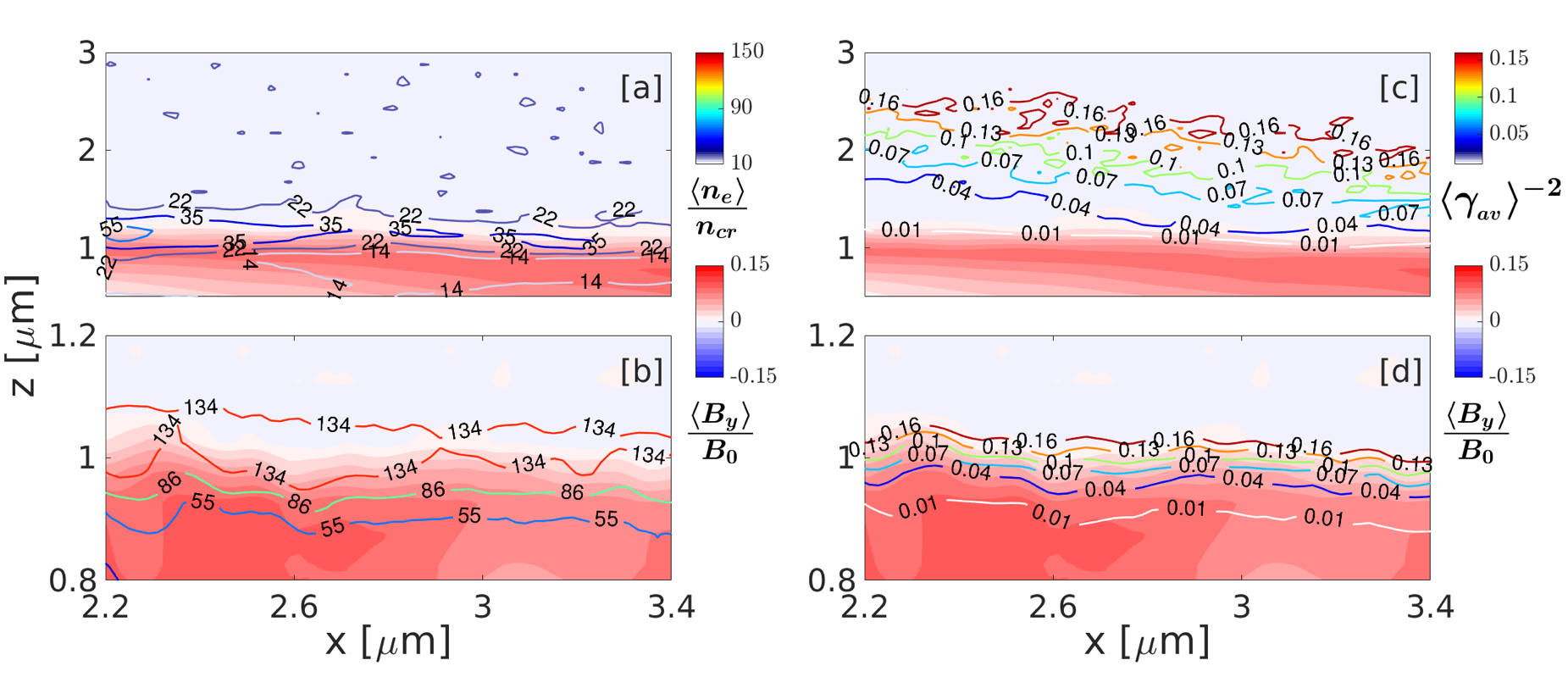}
    \caption{Contours of electron density $\langle{n_e}\rangle/{n_{cr}}$ (a and b) and $\langle\gamma_{av}\rangle^{-2}$ (c and d) plotted on top time-averaged magnetic field $\langle{B_y}\rangle/{B_0}$ for a target with a channel (b and d) and without a channel (a and c).}
    \label{FR_zoomed_dens}
\end{figure*}

In order to gain insight into relative changes of the field, density, and the $\gamma$-factor, we have examined the rotation caused by a thin plasma slice. The slice is only 0.05 $\mu$m or one cell thick in the $y$-direction, which eliminates integration along $y$. We chose the slice located at $y = 0$ $\mu$m primarily because the plasma magnetic field is almost perpendicular to this slice in the entire $(x,z)$-plane and thus its effect on the polarization rotation is maximized. We found that the maximum polarization rotation is almost 50 times higher for the structured target than for the initially uniform target.  

Figure \ref{FR_dependence} shows the correlation between the rotation angle $\phi$ and the local values of $\langle \gamma_{av} \rangle$, $\langle n_e \rangle$, and $\langle B_y \rangle$ that cause this rotation. We only show the top 10\% of $\phi$ from the $(x,z)$-plane, with each marker representing a separate point of the $(x,z)$-grid. The strongest signal in the structured target originates from a region where $\langle \gamma_{av} \rangle$ is roughly two times lower than $\langle \gamma_{av} \rangle$ of the initially uniform target. More importantly, the corresponding density in the structured target is significantly higher than the corresponding density in the initially uniform target. These values even exceed the initial density in the bulk of the structured target ($100 n_{cr}$), which suggests that the strong magnetic field is not fully contained in the channel.

All of the data-points shown in Fig. \ref{FR_dependence} are located at the periphery of the magnetic filament, because $\langle \gamma_{av} \rangle$ peaks at the center of the filament ($z = 0$) and decreases radially outwards. Figure \ref{FR_zoomed_dens} shows contours of $\langle n_e \rangle$ and $\langle \gamma_{av} \rangle^{-2}$ in a small region at the periphery of the magnetic filaments for the two types of targets. The background color is the strength of $\langle B_y \rangle$. We have plotted $\langle \gamma_{av} \rangle^{-2}$ rather than $\langle \gamma_{av} \rangle$, because this is the quantity that determines the polarization rotation for relativistic electrons. It is evident from Fig.~\ref{FR_zoomed_dens}b that the strong magnetic field of the structured target exists in a region that is much denser than the channel (see Fig.~\ref{FR_zoomed_dens}a). In this region, hot electrons from the channel mix with the cold electrons from the bulk, which causes a drop on $\langle \gamma_{av} \rangle$. As a result, the values of $\langle \gamma_{av} \rangle^{-2}$ are increased (compare Figs.~\ref{FR_zoomed_dens}c and \ref{FR_zoomed_dens}d) and the polarization rotation is increased as well compared to the case of the initially uniform target.

We then conclude that the structured target has two important benefits that lead to an increased polarization rotation: the bulk electrons increase the density in the region with a strong magnetic field, while simultaneously they reduce the characteristic $\gamma$-factor.

\begin{table}[H]
\centering
\begin{tabular}{| l | r | r | r | r | r | r |}
\hline
\bf f/\# & f1 & f2 & f3 & f4 & f5 & f6\\ \hline
  &  &  &  &  &  & \\
$\bf{I_L [W/cm^2]}$ &8.0$\times 10^{22}$&1.4$\times 10^{22}$&8.0$\times 10^{21}$&4.0$\times 10^{21}$&2.0$\times 10^{21}$&1.0$\times 10^{21}$\\ 
$\bf{\textit{\textbf{w}}_0 [\mu m]}$&0.50&1.20&1.58&2.24&3.16&4.47\\
$\bf{R_{opt}[\mu m]}$&0.8&1.2&1.3&1.3&1.7&1.5\\ \hline
\end{tabular}
\caption{Optimal channel radius $R_{opt}$ for different focal laser parameters, where the laser spot size $w_0$ is the FWHM for the laser intensity in the focal plane located at the entrance into the channel and $I_L$ is the laser peak intensity in the focal plane (in the absence of the target).}
\label{table_laser}
\end{table}


\section{Setup optimization for enhanced Faraday rotation}

As discussed above, we have already established that using a structured target is advantageous for the detection of strong laser-driven magnetic fields in a relativistically transparent plasma. In this section, we show that the setup can be further optimized by adjusting the focusing of the laser pulse and the radius of the channel, while keeping the laser pulse duration and energy fixed. In order to make our results experimentally relevant, we perform our optimization analysis for a 300 TW laser pulse that delivers 10 J of energy on target over 30 fs. These are the same laser parameters that are expected to be available at the  HED station at the European XFEL  where a 300 TW class Ti:Sa laser system is currently being installed by the HiBEF user consortium. 

Laser peak intensity is the key parameter that determines whether the irradiated target becomes relativistically transparent or not. The intensity can be adjusted by adjusting the size of the laser focal spot. In practice, this is usually achieved with the help of focusing optics and without having to adjust the duration or energy of the laser pulse. We use the same approach in our simulations to optimize the setup, where we vary the size of the focal spot $w_0$ to achieve six different peak laser intensities $I_L$ given in Table \ref{table_laser}. 

\begin{figure}[H]
\subcaptionbox*{}
[0.3\linewidth]
{\includegraphics[scale=0.3]{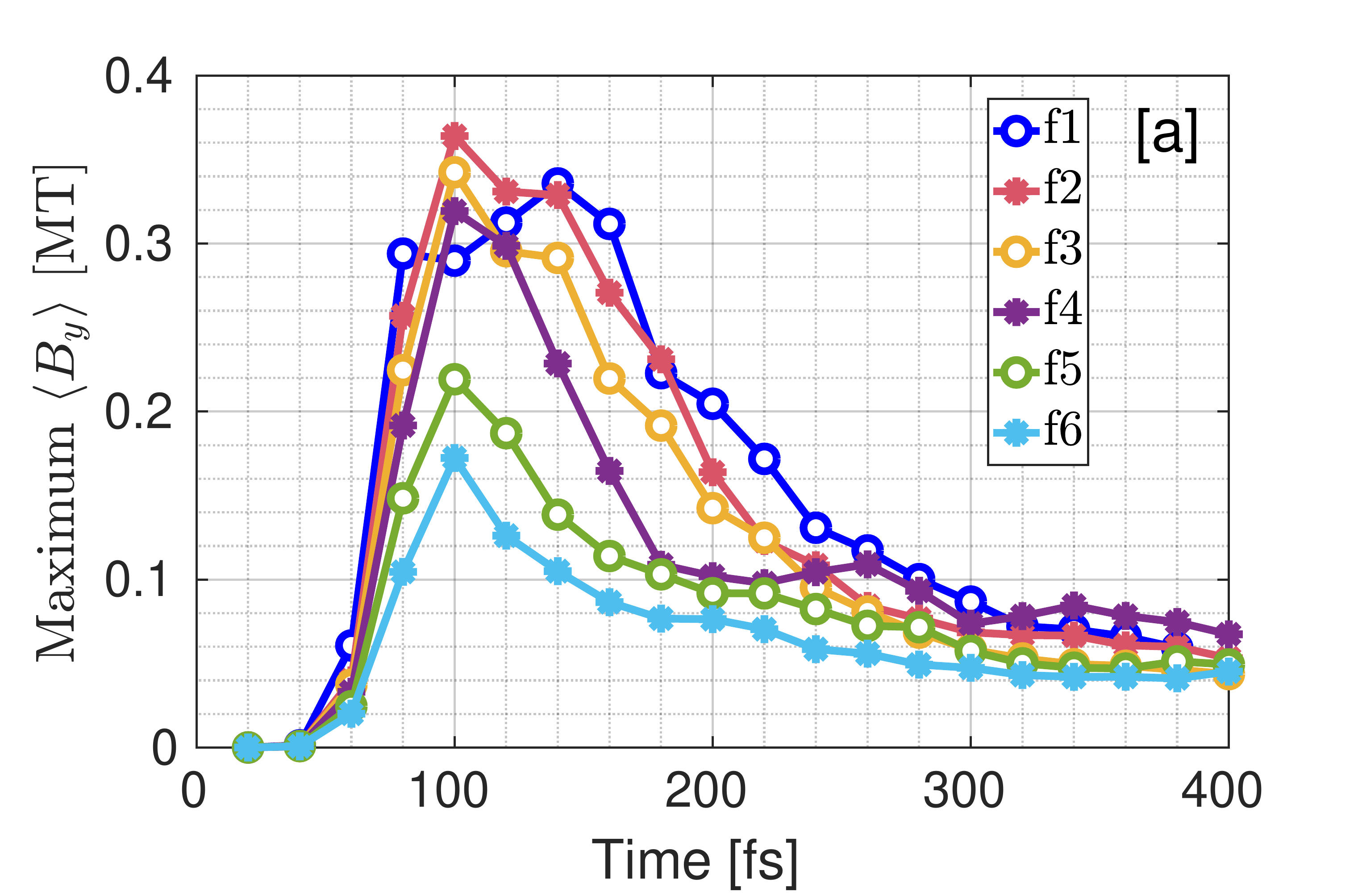}}
\hspace{40mm}
\subcaptionbox*{}
[0.4\linewidth]
{\includegraphics[scale=0.38]{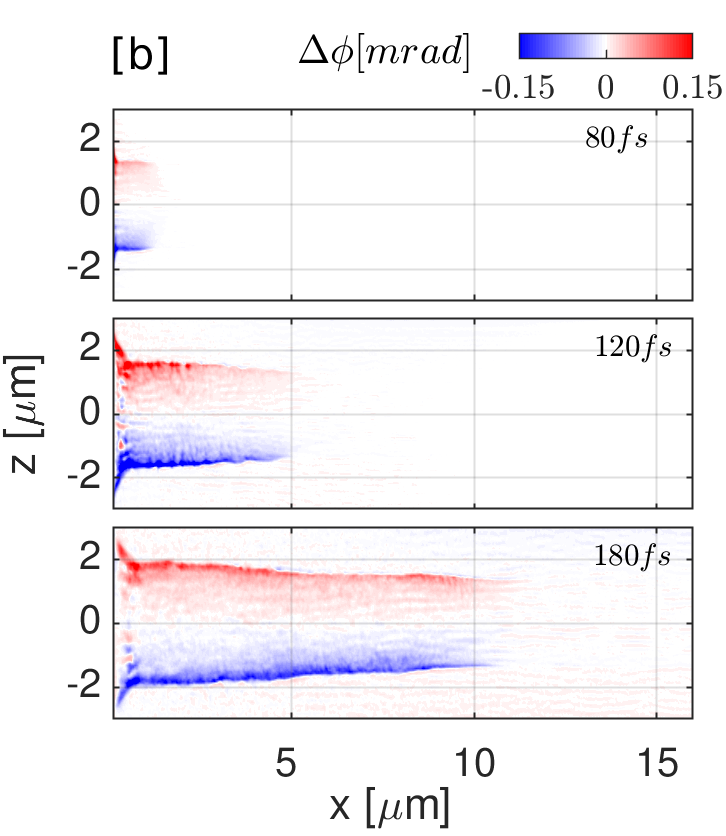}}
\caption{[a] The maximum strength of the laser-driven azimuthal magnetic field for six different laser parameters listed in Table \ref{table_laser}. [b] Three snapshots of the x-ray beam polarization rotation in the case of the f3 laser focusing configuration (see Table \ref{table_laser}).}
\label{RA_f3}
\end{figure}

There are additional reasons for performing the optimization. The tightest possible focusing that we have used so far in our simulations corresponds to an experimental setup with an f-number equal to 1, so this regime is difficult to achieve experimentally because of the associated stringent demands on the spatial phase of the laser beam. Moreover, the corresponding focusing optics has to be placed relatively close to the target, which increases the risk of damaging it during shots. Therefore, it would be preferable to use a setup with a higher f-number, provided that it can deliver a signal comparable to what we have calculated in the previous section. Furthermore, the laser defocusing can be beneficial since the azimuthal magnetic field linearly increases with the radius of the magnetic filament for a given longitudinal electron current density in the plasma. Results of additional simulations with higher f-numbers corroborate this statement. As shown in Fig.~\ref{RA_f3}a, the maximum azimuthal magnetic field indeed remains strong as we defocus the incoming laser pulse.

Figure~\ref{RA_f3}b shows snapshots of the x-ray polarization rotation induced by a target irradiated in an f3 focusing setup (see Table \ref{table_laser}). The radius of the channel is 1.3 $\mu$m. Even though the peak laser intensity is only $8 \times 10^{21}$ W/cm$^{2}$, the maximum rotation is comparable to that shown in Fig.~\ref{FR_w_wo_channel} for the f1 setup where the peak intensity is ten times higher. This indicates that there is indeed room for optimization. The maximum rotation angle is an important indicator of the magnetic field strength, but it may not be the best metric for comparing different focusing configurations because, strictly speaking, it only represents a single cell in the $(x,z)$-plane.

In order to compare different focusing configurations, we compute what we call the detectable area. This is the area in the $(x,z)$-plane where the polarization rotation of the probing x-ray beam exceeds the minimum detectable angle. Based on previously published results for the European XFEL setup~\cite{Huang2017}, we choose 10 $\mu$rad to be the threshold value. The detectable area is calculated for snapshots of the required plasma parameters that are averaged over four laser periods. On the other hand, an x-ray photon travels four laser wavelengths or 3.2 $\mu$m during this time interval. This distance is roughly equal to the diameter of the considered channel and thus our approach of using a single snapshot of time-averaged parameters is self-consistent\footnote{This approach can still be used for much thicker channels, but only if the time-averaged values evolve slowly on a time scale comparable to the travel time of an x-ray photon across the channel. }.

\begin{figure}[H]
\subcaptionbox*{}
[0.3\linewidth]
{\includegraphics[scale=0.26]{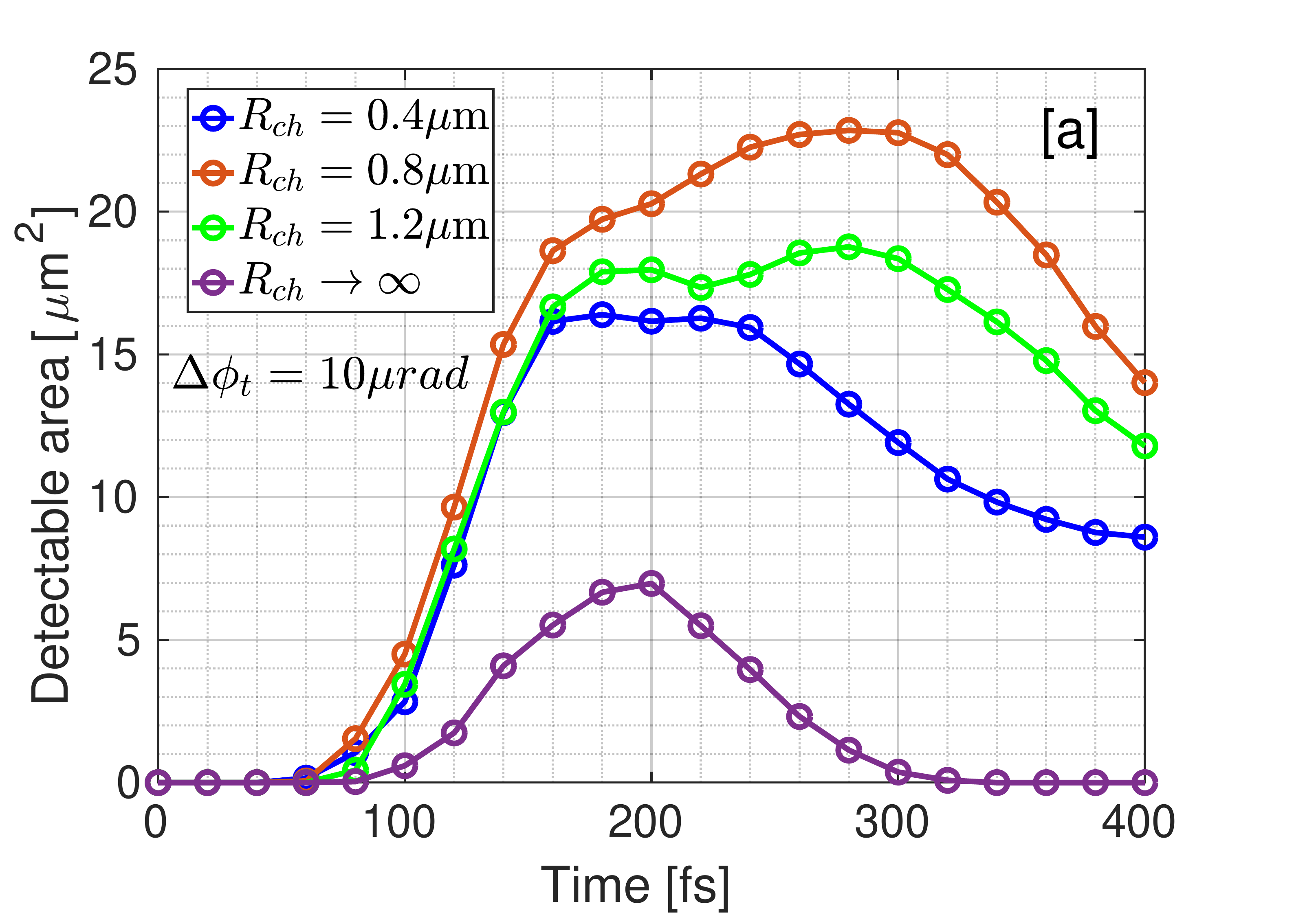}}
\hspace{30mm}
\subcaptionbox*{}
[0.3\linewidth]
{\includegraphics[scale=0.26]{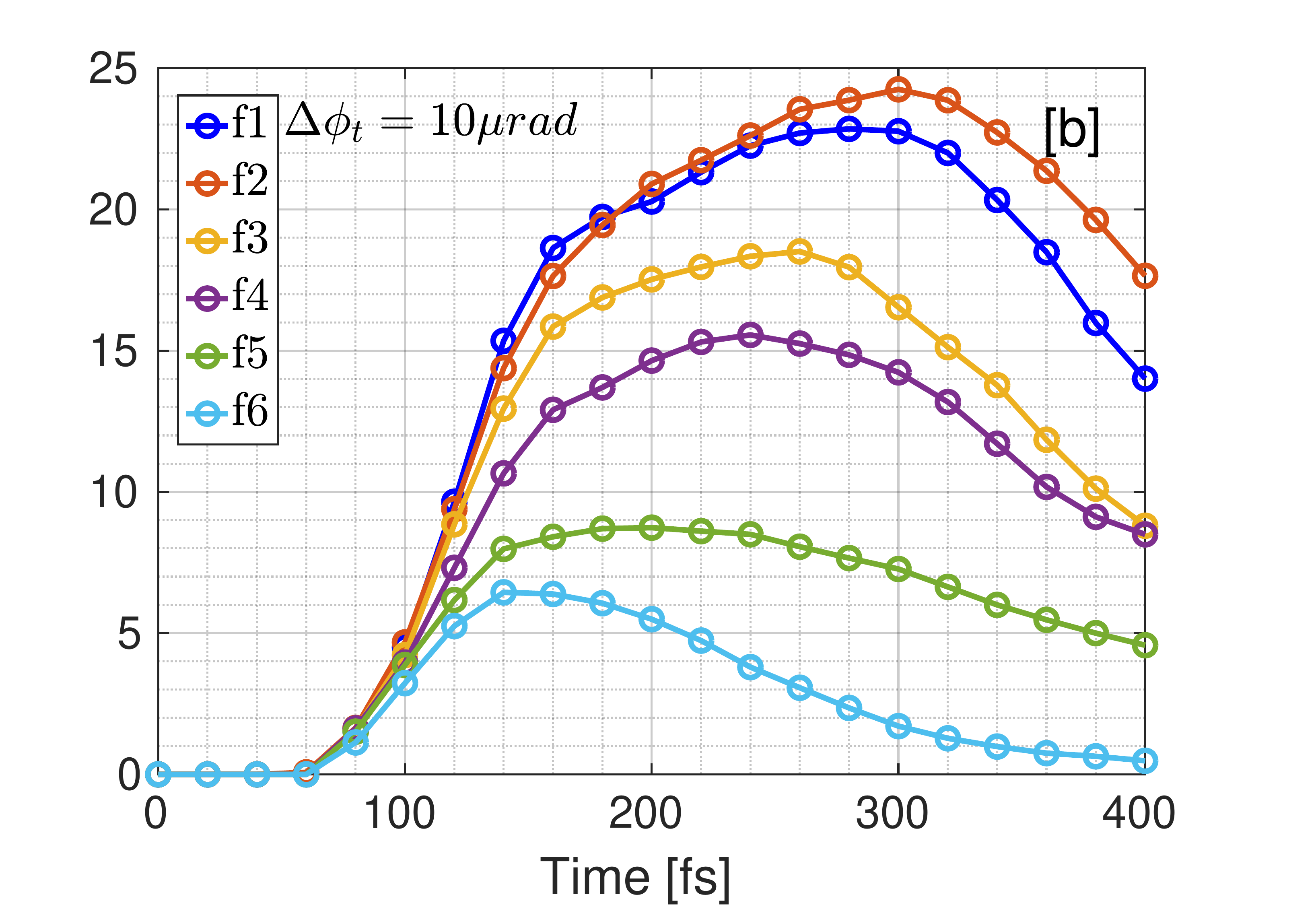}}
\caption{The two-step optimization of the detectable area with a threshold rotation angle of $\Delta \phi_t = 10$ $\mu$rad. (a) Optimization of the channel radius $R_{ch}$ for the f1 setup. The optimal radius is $R_{opt} = 0.8$ $\mu$m (see Table~\ref{table_laser}). (b) Optimization of the detectable area for different focusing setups with their own optimal radius $R_{opt}$ listed in Table~\ref{table_laser}.}
\label{RM_f1}
\end{figure}

We perform the optimization of the detectable area in two steps. Our first step is to optimize the channel radius for each laser configuration. Our second step is to compare the optimized detection areas for different focusing configurations. Figure~\ref{RM_f1}a shows how the detectable area changes in time for the f1 setup and four different target configurations. In this case, the optimal channel radius is $R_{opt} \approx 0.8$ $\mu$m. This is the target configuration that was used in the simulations discussed in Section~\ref{Sec-5} (see Table~\ref{table_PIC} for more details). Employing the same approach, we found $R_{opt}$ for all of the considered laser focusing configurations, with the results of the optimization listed in Table~\ref{table_laser}. The optimal channel radius is essentially the same for f2, f3, and f4 setups. This nontrivial outcome underscores the importance of the adopted optimization approach.

Figure~\ref{RM_f1}b shows how the detectable area changes in time for each of the considered laser focusing configurations with a corresponding optimal channel radius $R_{opt}$. The biggest detectable area is achieved by using the f2 setup rather than the f1 setup that delivers the highest on-target laser intensity. These results indicate that there is flexibility in terms of the focusing setups. Our main conclusion is that choosing an f2 setup for the first day experiments at the HED instrument at the European XFEL would allow us to meet stringent demands on the experimental setup and laser beam focusing quality without reducing the rotation or the magnetic field strength.

\section{\sffamily{Synthetic diagnostic for Faraday rotation of an x-ray probe beam}}

In this section we use our already discussed 3D simulation results to calculate a simplified diagnostic readout in order to determine how many photons we can realistically expect in a single shot and whether multiple shots are needed.

Our calculation are guided by the experimental capabilities of the European XFEL that can deliver 6.457 keV x-ray beams of variable duration. We consider two scenarios. In the first scenario, the x-ray beam duration is 10 fs and the number of x-ray photons is $5 \times 10^{11}$. In the second scenario, the x-ray beam duration is 100 fs and the number of x-ray photons is ten times higher. The x-ray beam cross section that we denote as $S_{probe}$ is 20 $\mu$m by 20 $\mu$m in both cases, which yields $\delta N_{probe} \approx N_{probe} / S_{probe}$ x-ray photons per unit area of the target. We use this photon number to probe the magnetic field in our 3D PIC simulations. The final result can be easily re-scaled if this number is different.

\begin{figure}[htbp!]
	\centering
	\includegraphics[scale=0.52,trim={.8cm 1.8cm 1.2cm 1cm},clip]{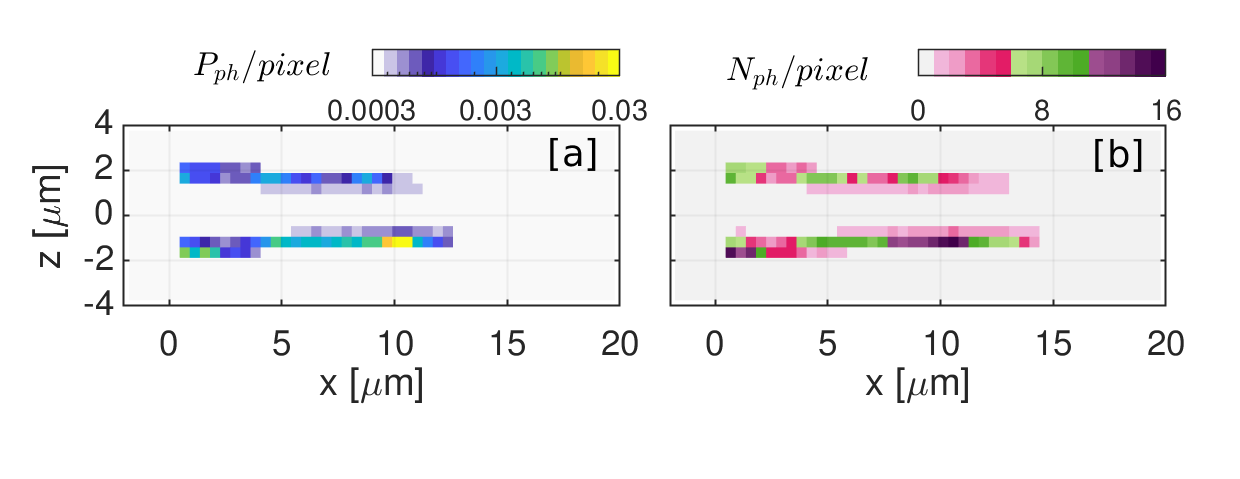}
\caption{Predicted signal for 10 fs and 100 fs x-ray beams probing the magnetic field filament generated using the optimized setup for the f2 laser focusing configuration (see Table~\ref{table_laser}). (a) The probability for a photon to hit a single CCD camera pixel, as defined by Eq.~(\ref{probability_per_pixel}). The 10 fs beam probes the magnetic filament over 165 fs $ \leq t \leq $ 175 fs. (b) The number of photons per pixel, as defined by Eq.~(\ref{photons_per_pixel}), for a 100 fs beam probing the magnetic filament over a time interval 140 fs $\leq t \leq $ 240 fs.}
    \label{RA_sum_f1}
\end{figure}

We have already shown that the polarization plane of the x-ray beam rotates after it traverses the magnetic field filament. Let us assume that the original x-ray beam is linearly polarized along the $x$-axis. Then the transmitted beam has a $z$-polarized component whose amplitude is $E_z = E_0 \sin (\Delta \phi) \approx E_0 \Delta \phi$, where $E_0$ is the amplitude of the incoming x-ray beam and $\Delta \phi$ is a very small polarization rotation angle. Here we neglect any x-ray absorption. If this transversely polarized component can be separated from the transmitted beam, then the intensity of the resulting beam would be smaller than the intensity of the original probe beam by a factor of $\left( \Delta \phi \right)^2$. In practice, this can be achieved by employing channel cut crystals~\cite{MARX2011915,marx2013_xray_polarimetry}. In order to make our results easier to interpret, we neglect any loss of photons representing the rotated (transversely polarized) component. In reality, this is not necessarily the case and our results would need to be re-scaled using an appropriate multiplier representing the photon loss for a given detection setup.

Using the described approach, we find that the number of photons per unit area in the transversely polarized beam is 
\begin{equation}
    \delta N_{out} = \left( \Delta \phi \right)^2 \delta N_{probe} = \left( \Delta \phi \right)^2 N_{probe} / S_{probe},
\end{equation}
where $\left( \Delta \phi \right)^2$ is a function of $x$ and $z$. A photon detector has its own spatial resolution that we would like to take into account in our analysis because it is likely to impact the resolution of the detected signal. In the case of a CCD camera~\cite{Huang2017}, the resolution is dictated by the size of a pixel. We find that, for a magnification factor of 30, a pixel with transverse dimension of 13 $\mu$m by 13 $\mu$m  corresponds to an area of $S_{pixel} \approx $ 0.2 $\mu$m$^2$ in the cross section of the target [i.e. an area in the $(x,z)$-plane of the simulation]. The corresponding dimensions are 0.43 $\mu$m by 0.43 $\mu$m. We use the following expression to compute the number of photons per pixel using our simulation results
\begin{equation} \label{photons_per_pixel}
    N_{ph} / pixel \equiv \int_{S_{pixel}} \delta N_{out} dx dy = N_{probe} \int_{S_{pixel}} \left( \Delta \phi \right)^2 \frac{dx dy}{S_{probe}}.
\end{equation}
Anticipating that the total photon count, $N_{out}$ is likely to be low, we also introduce a probability for a photon to hit a given pixel
\begin{equation} \label{probability_per_pixel}
    P_{ph} / pixel \equiv \frac{N_{ph} / pixel}{N_{out}}.
\end{equation}
This probability can be used to generate the detector signal for a limited number of detected photons using a Monte Carlo algorithm. Equations~(\ref{photons_per_pixel}) and (\ref{probability_per_pixel}) are provided in order to improve the usability of our results by making it clear that using a different pixel size ($S_*$ in the $(x,y)$-plane) requires one to simply multiply our results by a ratio of $S_* / S_{pixel}$. 

Figure~\ref{RA_sum_f1}a shows the discussed probability for a 10 fs x-ray beam probing the magnetic field filament generated using the optimized setup for the f2 laser focusing configuration (see Table~\ref{table_laser}). As expected, the total number of detected photons is very low in this case, $N_{out} \approx 76$. It takes roughly 12 fs for the x-ray photons to traverse the magnetic filament. We use a single snapshot taken at $t = 170$ fs of time-averaged quantities from our 3D simulation to obtain the result shown in Fig.~\ref{RA_sum_f1}a, with the time-averaging performed over four laser periods.

Using a longer probe pulse of 100 fs roughly increases the number of detected photons by an order of magnitude to $N_{out} \approx 670 $. The magnetic filament evolves on this time scale, so the calculation is performed using five different snapshots of time-averaged quantities that are taken 20 fs apart. The result is shown in Figure~\ref{RA_sum_f1}b, where we plot the number of photons per pixel, $N_{ph} / pixel$. The pixels with less than one photon are not shown on this plot. It appears that the total number of detected photons is sufficient to resolve the shape of the magnetic filament embedded in a target. 

We conclude this section by discussing the impact of the photon polarization purity in the probe beam. We have so far assumed that the incoming beam is linearly polarized along the $x$-axis. In reality, the probe beam might contain a small number of photons that are already polarized along the $z$-axis. The polarization purity $\mathcal{P}$ is usually defined as a ratio of the number of these photons to the total number of photons~\cite{marx2013_xray_polarimetry}. It is the same as the ratio of the intensity of the $z$-polarized part of the beam to the total beam intensity. These photons would contaminate the signal that we have calculated and plotted in Fig.~\ref{RA_sum_f1} by introducing noise. The strength of our signal is approximately
\begin{equation} 
    N_{ph} / pixel \approx N_{probe} \left( \Delta \phi \right)^2 \frac{S_{pixel}}{S_{probe}},
\end{equation}
whereas the expected noise level due to the finite polarization purity $\mathcal{P}$ is
\begin{equation} 
    N_{ph}^{noise} / pixel \approx N_{probe} \mathcal{P} \frac{S_{pixel}}{S_{probe}}.
\end{equation}
By comparing the two expressions, we conclude that the noise level might be tolerable if the polarization purity satisfies the requirement
\begin{equation} \label{purity_criterion}
    \mathcal{P} \ll \left( \Delta \phi \right)^2
\end{equation}
for a characteristic rotation angle predicted by the simulations. In our case, we have $\Delta \phi \approx 10^{-4}$ rad, which sets a maximum acceptable value for the polarization purity of $\mathcal{P}_{\max} \approx 10^{-8}$. The polarization purity for the European XFEL is expected satisfy the condition  (\ref{purity_criterion}) by employing a polarimetry setup similar to that discussed in Ref.~\cite{marx2013_xray_polarimetry}.

The results presented in this section of the manuscript should be viewed as an upper estimate for the number of detectable photons, denoted as $N_{out}$. An experimental detection setup would necessarily have a transmission coefficient $\cal{T}$ that is less than unity, which would reduce the number of photons to ${\cal{T}} N_{out}$. We anticipate a value of ${\cal{T}} \approx 10^{-2}$. In this case, the number of photons would become too low for single-shot detection and the magnetic field detection would require accumulating the signal over at least tens of shots.

\section{\sffamily{Summary and conclusions}}

We have considered a setup where a laser pulse of extreme intensity drives a strong quasi-static azimuthal magnetic field in a classically over-critical plasma with $n_e \gg n_{cr}$. This paper examines the feasibility of using an x-ray beam from the European XFEL for the detection of the magnetic field. The magnetic field embedded in the plasma rotates the polarization plane of the incoming x-ray beam. We find that for the 6.457 keV x-ray probe beam and a 0.4 MT magnetic field the strongest rotation occurs when the x-ray beam propagates parallel to the magnetic field lines, which is the Faraday rotation (see Appendix~\ref{App_A}).

Relativistically induced transparency plays a critical role in generating the considered volumetric magnetic field, as it allows the laser pulse to propagate through the over-critical plasma. However, our simulations show that the relativistically induced transparency also significantly reduces the Faraday rotation and it is therefore detrimental in the context of the considered probing technique. We have shown, for the first time, that the rotation angle can be increased by roughly an order of magnitude by employing structured targets that contain a relativistically transparent channel surrounded by relativistically near-critical material. The enhancement takes place because the plasma density increases at the periphery where the magnetic field is strong, while, at the same time, the local average $\gamma$-factor drops due to mixing between hot electrons from the channel with cold electrons from the surrounding material. The resulting rotation angle is roughly $\Delta \phi \approx 10^{-4}$ rad.

We have also considered different laser focusing configurations while keeping the energy of the incoming laser pulse in our 3D PIC simulations constant. Our main conclusion is that choosing an f2 setup for the first day experiments at the HED instrument at the European XFEL would allow us to meet stringent demands on the experimental setup and laser beam focusing quality without reducing the rotation or the magnetic field strength. The Faraday rotation produces an x-ray beam that is transversely polarized with respect to the original probe beam. The detection of this beam is a possible path towards the detection of the strong magnetic field embedded in a dense plasma. We have evaluated the number of photons in this beam and found that it is roughly $10^3$ for a 100 fs long probe beam with $5 \times 10^{12}$ photons. We have concluded that based on the predicted rotation angle the polarization purity must be much better than $\mathcal{P} \approx 10^{-8}$ in order to detect the signal above the noise level. We also find that single-shot detection may not be feasible if the transmission coefficient of the experimental detection setup is $10^{-2}$. In this case, the photon signal must be accumulated of tens of shots.  

We have also performed additional 3D PIC simulations to determine what one should expect for a different channel material and a different channel density (see Appendix~\ref{App_B}). Even though all of the simulations are performed for a hydrogen-filled channel with $n_e = 20 n_{cr}$, we find that another material can be used to fill the channel without affecting the outcome if that is more practical in terms of target manufacturing techniques. A carbon-filled channel produces an almost identical magnetic filament. However, we find that the electron density in the channel is an influential parameter. A denser channel ($n_e = 40 n_{cr}$) slows down the laser pulse propagation and the magnetic field filament becomes shorter while the maximum rotation angle remains unaffected. This can reduce the detectable area, causing an appreciable reduction in the number of the detectable x-ray photons. A detailed parameter scan is required to determine the optimal electron density in the channel.



\section*{\sffamily{Acknowledgements}}

This research was supported by the Air Force Office of Scientific Research under Grant No. FA9550-17-1-0382 and the National Science Foundation under Grant No. 1632777. Particle-in-cell simulations were performed using EPOCH~\cite{Epoch}, developed under UK EPSRC grants EP/G054940, EP/G055165, and EP/G056803. High performance computing resources were provided by Texas Advanced Computing Center (TACC) at The University of Texas at Austin and by the Extreme Science and Engineering Discovery Environment (XSEDE) through allocation \textsl{PHY180033}. Data collaboration supported by the SeedMe2 project~\cite{Seedme2} (http://dibbs.seedme.org).


\appendix

\section*{\sffamily{Appendices}}
\section{\sffamily{Reduction of the Faraday and Cotton-Mouton effects\\in a relativistic plasma}} \label{App_A}

It is well-known that the optical properties of a plasma are dependent on the average electron energy if this energy is relativistic. The Faraday rotation is no exception, but the impact of electron heating on the polarization rotation is often characterized using electron temperature~\cite{FR_rel2,FR_rel3,FR_rel4,FR_Trubnikov_phd,FR_melrose_1997}. This is not a convenient parameter for our present work, because the distribution function is not an input function. A more convenient parameter is the distribution-averaged relativistic $\gamma$-factor, $\gamma_{av}$, since this parameter can be readily computed in a kinetic simulation. 

The polarization rotation is given by Eq.~(\ref{brot_correct}), where $\alpha$ represents the change of the optical properties caused by relativistic electron motion in a uniformly heated plasma with a uniform density. In what follows, we determine the expression for $\alpha$ in terms of $\gamma_{av}$ that is then used in the main part of the paper to evaluate the polarization rotation in a non-uniformly heated plasma with density gradients. Our approach is to perform a parameter scan where the Faraday rotation is computed for different initial values of $\gamma_{av}$. The plasma is irradiated by a low amplitude plane electromagnetic wave, so that the electron heating by the wave is negligible and the wave can be treated as a probe. 

We have performed a set of one-dimensional (1D) PIC simulations where the electron distribution is initialized as an isotropic Maxwell-J\"uttner distribution in momentum space~\cite{juttner1911maxwellsche,rezzolla2013relativistic},
\begin{equation} \label{MJ}
f(p)dp = \dfrac{n_e}{\zeta K_2(\zeta^{-1})} \textrm{exp} \left( - \frac{1}{\zeta} \sqrt{1+\dfrac{p^2}{m_e^2c^2}} \right) \dfrac{p^2 dp}{\left( m_e c \right)^3}.
\end{equation}
Here $p$ is the amplitude of the electron momentum, $n_e$ is the electron density, $K_2$ is the modified Bessel function of the second kind, and 
\begin{equation}
    \zeta \equiv T_e \left/ m_e c^2 \right.
\end{equation}
is the normalized electron temperature. The distribution given by Eq.~(\ref{MJ}) reduces to a non-relativistic Maxwellian distribution for $\zeta \ll 1$. The original version of EPOCH~\cite{Epoch} was modified in order to load macro-particles according to the Maxwell-J\"uttner momentum distribution. In the modified version, the macro-particles are loaded using the Sobol method detailed by S. Zenitani~\cite{zenitani2015loading} that employs rejection sampling and spherical scattering.

\begin{figure}[H]
\subcaptionbox*{}
[0.4\linewidth]
	{\includegraphics[width=0.5\textwidth,trim={0.0cm 0.0cm 0 0.0cm},clip]{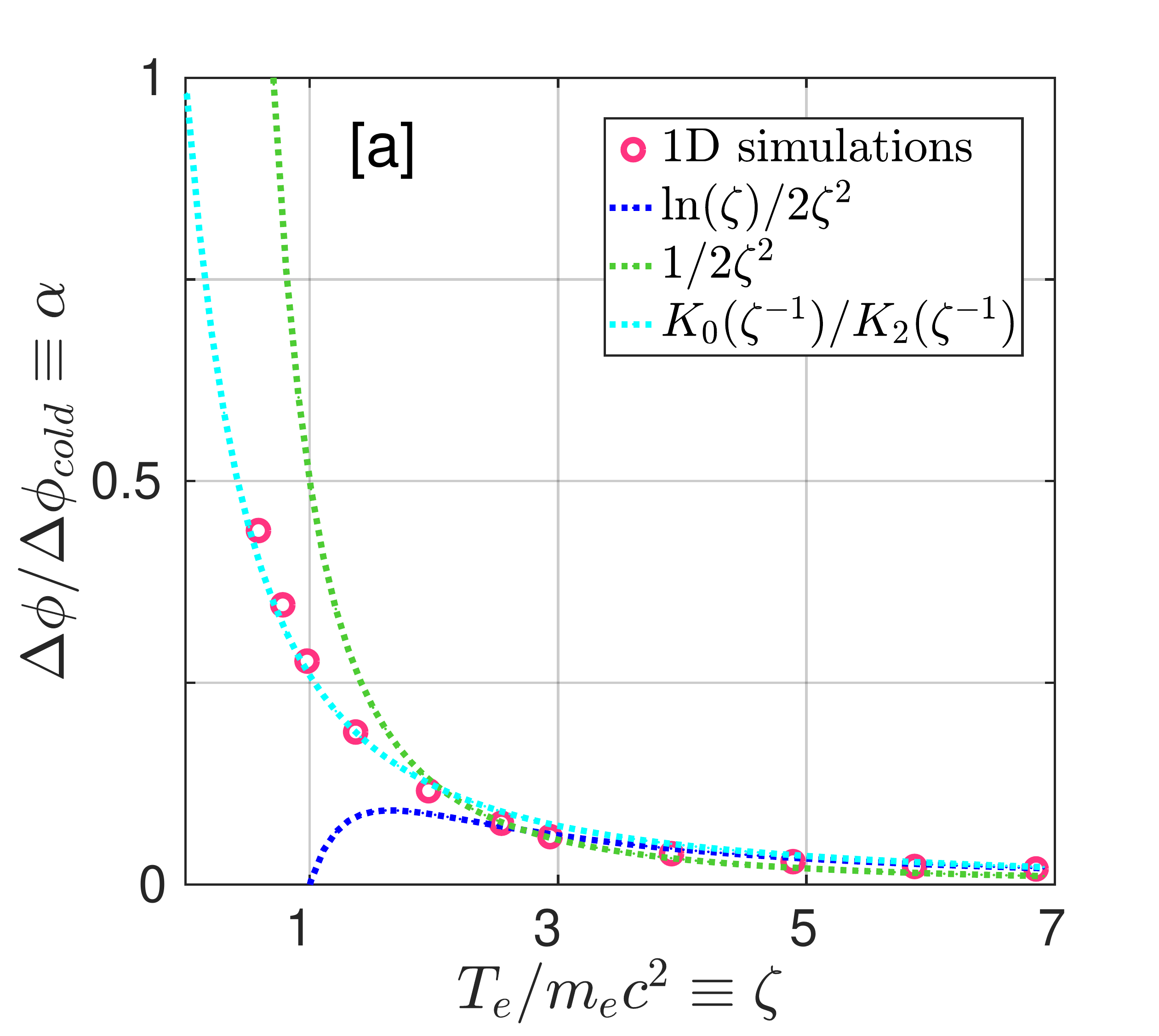}}
    \hspace{15mm}
\subcaptionbox*{}
[0.4\linewidth]
	{\includegraphics[width=0.5\textwidth,trim={0.0cm 0.0cm 0 0.0cm},clip]{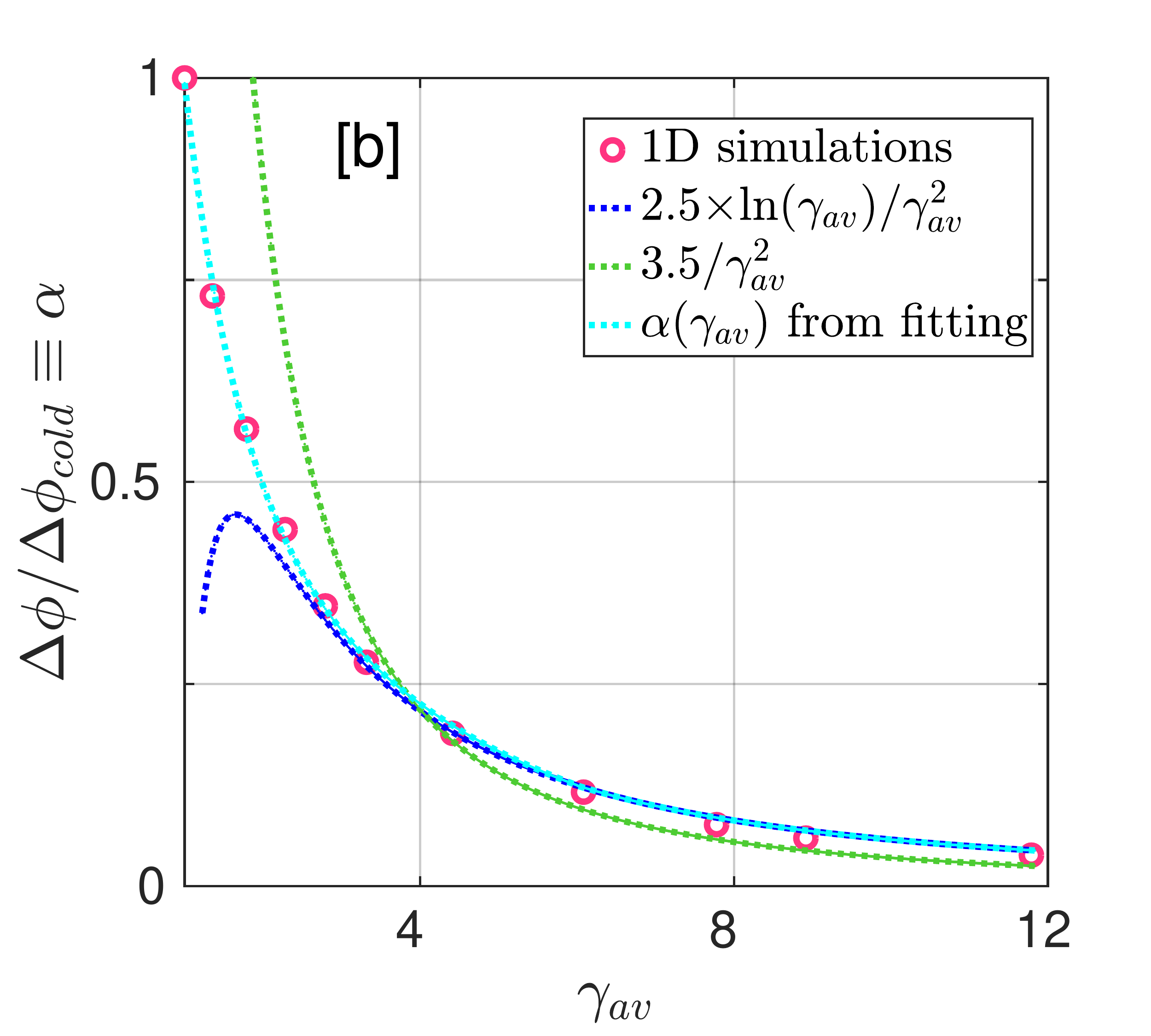}}
\caption{Relativistic correction to the Faraday rotation, $\alpha = \Delta \phi$/$\Delta \phi_{cold}$, as a function of electron temperature $T_e$ (a) and average relativistic $\gamma$-factor $\gamma_{av}$ (b). The circles are the results of an electron temperature scan obtained by performing 1D PIC simulations for a Maxwell-J\"uttner distribution given by Eq.~(\ref{MJ}) with different values of $\zeta = T_e / m_e c^2$. }
  \label{Dependence_FR_1D}
\end{figure}

In our 1D PIC simulations, a uniform plasma with $n_e = 7.2 \times 10^{25}$ m$^{-3}$ is irradiated by a linearly polarized probe pulse propagating along the $x$-axis. An externally applied 40 kT uniform magnetic field is also directed along the $x$-axis. The pulse amplitude increases from zero to $a_0 \approx 0.022$ and then remains constant. The electric field in the incoming pulse has only one component, which is $E_z$. Without any loss of generality, we use a pulse with 30 eV photons whose corresponding wavelength is 0.04 $\mu$m instead of using a 6.457 keV x-ray pulse whose wavelength is much shorter. As discussed in Section~\ref{Sec-3}, the polarization rotation scales as $\Delta \phi \propto 1 / \omega_*^2$, which is equivalent to $\Delta \phi \propto \lambda^2$, where $\lambda$ is the vacuum wavelength corresponding to the pulse frequency $\omega_*$. Using a pulse with a longer wavelength allows us to observe appreciable rotation over a much shorter distance, making the 1D scan shown in Fig.~\ref{Dependence_FR_1D} less demanding in terms of computational resources.

\begin{figure}[H]
	\centering
	\includegraphics[scale=0.45,trim={0cm .5cm 0.3 0cm},clip]{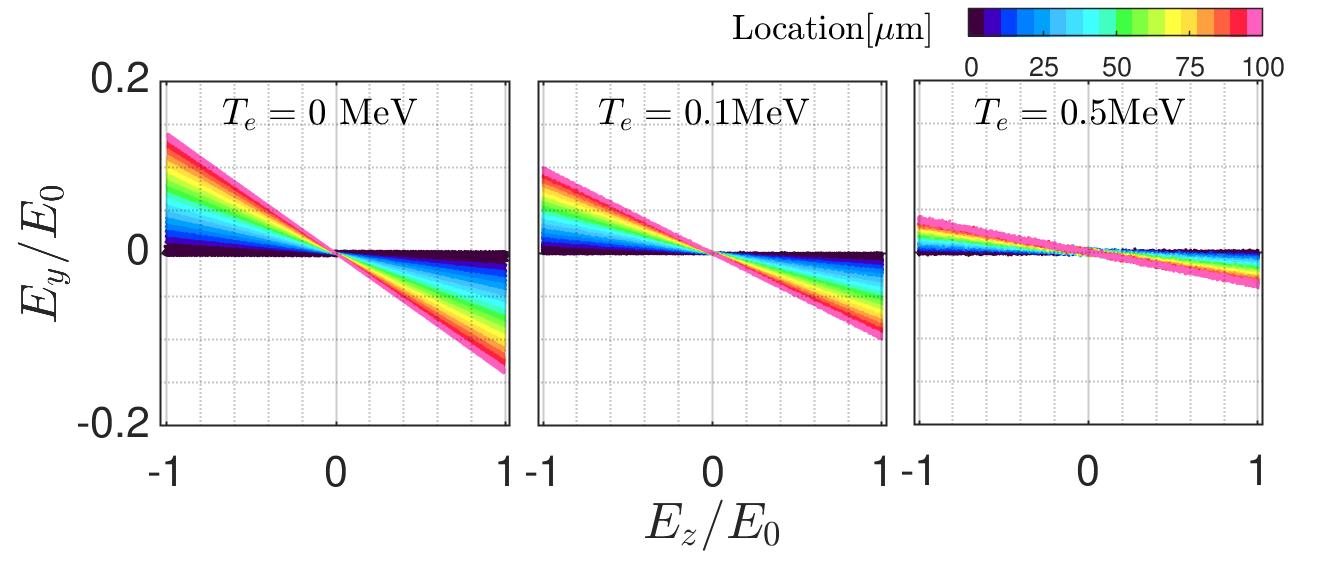}
	\caption{Spatially resolved Faraday rotation induced by a plasma with $T_e = 0$ MeV, $T_e = 0.1$ MeV, and $T_e = 0.5$ MeV. The color-coding indicates the distance along the magnetic field lines from the plasma edge where the pulse enters the plasma. Note that the data aspect ratio is set to 1:0.2 to clearly display the rotation.
	}
    \label{FR_1D_ex}
\end{figure}

Figure~\ref{FR_1D_ex} shows examples of the calculated Faraday rotation in our 1D PIC simulations for three different electron temperatures. The electric field remains essentially linearly polarized, with the biggest effect being the rotation of the polarization. The rotation angle increases linearly with the distance from the plasma edge, which is consistent with Eq.~(\ref{brot_correct}). The relativistic correction $\alpha$ is determined by dividing $\Delta \phi$ in a plasma with a given $T_e$ by $\Delta \phi$ in a plasma with cold electrons that we denote as $\Delta \phi_{cold}$. The polarization rotation $\Delta \phi$ is compared for the same distance from the plasma edge in both cases.

The result of a scan that yields $\alpha = \Delta \phi / \Delta \phi_{cold}$ as a function of $T_e$ is shown in Fig.~\ref{Dependence_FR_1D}a. The relativistic correction $\alpha$ is well described by ${K_0(\zeta^{-1})}/{K_2(\zeta^{-1})}$~\cite{FR_rel2,FR_rel3,FR_rel4} over a wide range of relativistic electron temperatures. At high temperatures, the correction is well approximated by $\ln (\zeta)/2 \zeta^2$. There is a one-to-one correspondence between $\zeta$ and $\gamma_{av}$ that we determine numerically for the Maxwell-J\"uttner distribution. We then use it to plot the scan result from Fig.~\ref{Dependence_FR_1D}a versus $\gamma_{av}$ in Fig.~\ref{Dependence_FR_1D}b. 

The relativistic correction $\alpha$ in Fig.~\ref{Dependence_FR_1D}b roughly scales as $1/\gamma_{av}^2$ for $\gamma_{av} \gg 1$. This trend agrees well with the estimate given by Eq.~(\ref{FR_hot}). As expected, our result also shows that there is no correction, i.e. $\alpha = 1$, in the limit of $\gamma_{av} \rightarrow 1$. In order to obtain an expression for $\alpha$ in the intermediate range, we used numerical fitting and found that the simulation results are well approximated by
\begin{equation}
\alpha = 
\begin{cases}
	2.141 \exp \left( -1.508  \gamma_{av} \right) + \ 0.6913 \exp \left( -0.2856 \gamma_{av} \right),\  \text{for } \gamma_{av} \leq 6\\
    2.5 \gamma_{av}^{-2} \ln \left( \gamma_{av} \right), \ \text{for } \gamma_{av} > 6.
\end{cases}
\label{eq_Xgamma}
\end{equation}

We have so far only considered the case where the probe beam propagates parallel to the magnetic field lines. However, the polarization of the incoming beam also changes if the beam propagates at an angle to the magnetic field lines. In what follows, we assess the impact of the magnetic field for a limiting case of perpendicular propagation.

If the plasma is cold, then the effect of the magnetic field is well-known. The polarization of the transverse electric field changes from linear to elliptical as the incoming electromagnetic wave enters and propagates through the plasma. This effect is often referred to as the Cotton-Mouton (CM) effect. In contrast to the parallel case, the propagating wave is now a superposition of an O-mode and an X-mode that are both linearly polarized in the plane transverse to the direction of the propagation. The phase velocity of these two modes are different, so their superposition becomes elliptically polarized as the phase difference, $\Delta \phi_{CM}$, accumulates. It should be pointed out that the effect disappears if the transverse electric field is either purely parallel or perpendicular to the magnetic field.

In the cold plasma limit, the phase difference between two eigenmodes for the propagation parallel and perpendicular to the magnetic field are given by~\cite{book_FR_CM_cold} 
\begin{eqnarray}
&& \Delta \phi_{\parallel}^{cold} \equiv  \Delta \phi_{FR}^{cold} = \frac{1}{2} \frac{\Delta l}{c} \frac{\omega_{pe}^2 \omega_{ce}}{\omega_{*}^2}, \label{FR_A} \\
&& \Delta \phi_{\perp}^{cold} \equiv \Delta \phi_{CM}^{cold} = \frac{1}{2} \frac{\Delta l}{c} \frac{\omega_{pe}^2 \omega_{ce}^2}{\omega_{*}^3} . \label{CM_A}
\end{eqnarray}
In the case of parallel propagation, it is the already discussed effect of the Faraday rotation (FR). Here $\omega_{pe}$ is the electron plasma frequency, $\omega_{ce}$ is the electron cyclotron frequency, $\omega_{*}$ is the frequency of the probe beam, and $\Delta l$ is the distance travelled by the probe beam. It follows from Eqs.~(\ref{FR_A}) and (\ref{CM_A}) that the ratio between the two is
\begin{equation}
     \frac{\Delta \phi_{\perp}^{cold}}{\Delta \phi_{\parallel}^{cold}} = \frac{\omega_{ce}}{\omega_*}.
\end{equation}
We find that 
\begin{equation} \label{ratio_exp}
\omega_{ce} / \omega_* \equiv \left( \omega_{ce} / \omega_* \right)_{exp} \approx 7 \times 10^{-3}   
\end{equation}
for a 6.457 keV x-ray probe beam and a 0.4 MT magnetic field. We therefore conclude that, for the experimentally relevant parameters, the phase shift is suppressed by at least two orders of magnitude if the probe beam travels perpendicular to the magnetic field in a cold plasma.

\begin{figure}[H]
	\centering
	\includegraphics[scale=0.35,trim={0cm .0cm 0.0cm 0cm},clip]{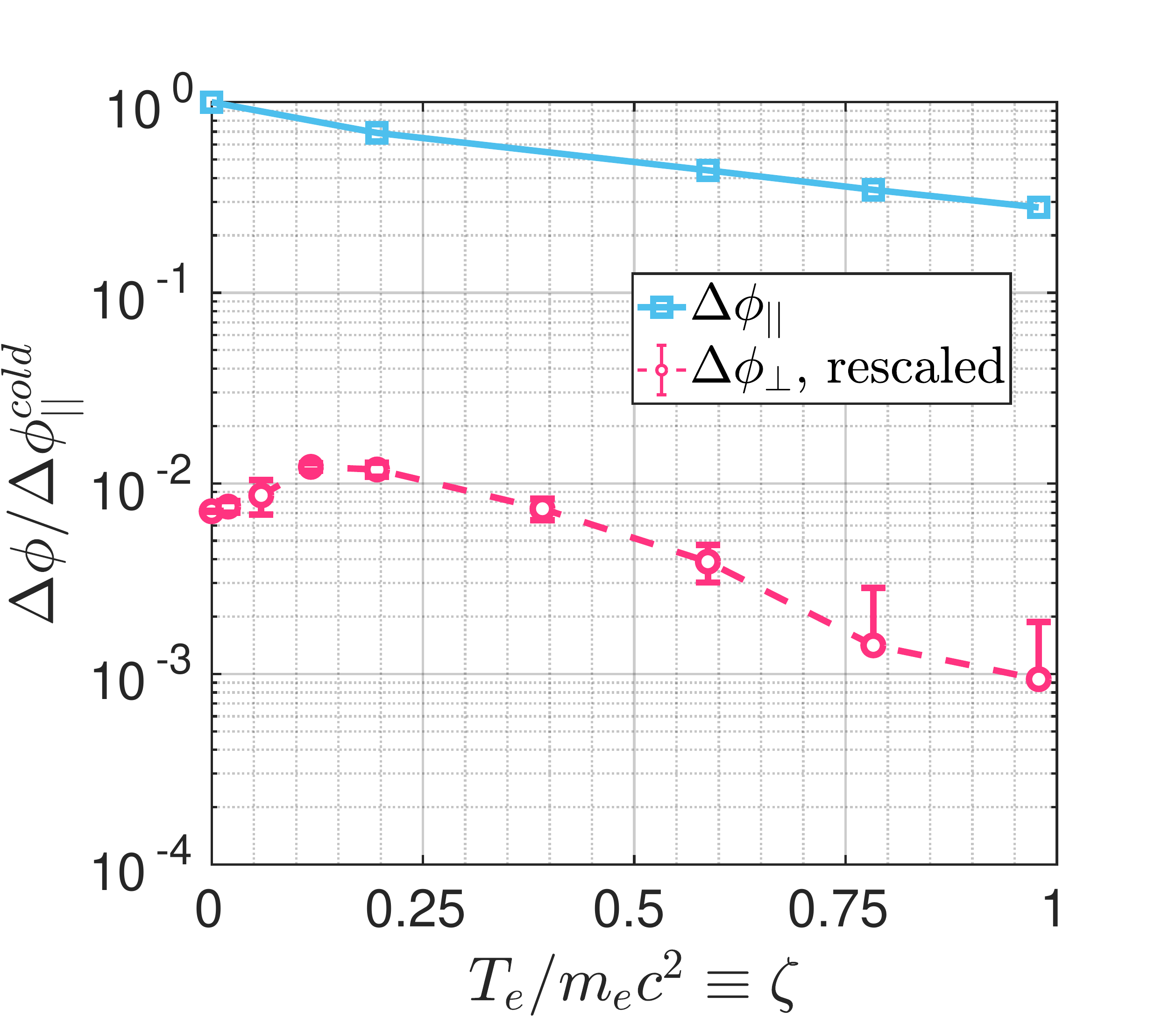}
	\caption{The phase shifts for longitudinal $(\Delta \phi_{\parallel})$ and transverse $(\Delta \phi_{\perp})$ propagation of a probe pulse in a plasma with relativistic electrons. Both quantities are normalized to the phase shift in a cold plasma $\Delta \phi_{\parallel}^{cold}$. The plotted values of $\Delta \phi_{\perp}$ are the numerically calculated values multiplied by $\left( \omega_{ce} / \omega_* \right)_{exp} / \left( \omega_{ce} / \omega_* \right)_{sim} \approx 4.7 \times 10^{-2}$.} 
    \label{CM_FR_1D}
\end{figure}

In order to confirm that $\Delta \phi_{\perp} / \Delta \phi_{\parallel}$ remains small at relativistic electron temperatures, we have performed a series of 1D simulations whose results are shown in Fig.~\ref{CM_FR_1D}. We use the same plasma and probe beam parameters that were used to obtain Figs.~\ref{Dependence_FR_1D}a and \ref{Dependence_FR_1D}b. In the case of perpendicular propagation, the polarization plane of the linearly polarized incoming beam is tilted by $\pi/4$ with respect to the magnetic field. This ensures that there is no significant difference between the amplitudes of the O and X-modes excited in the plasma, which makes the numerical detection of $\Delta \phi_{\perp}$ easier. In order to aid the comparison between $\Delta \phi_{\parallel}$ and $\Delta \phi_{\perp}$, we normalize the two quantities by $\Delta \phi_{\parallel}^{cold}$. Note that $\Delta \phi_{\parallel}^{cold}$ is the biggest possible phase difference over the same propagation length and it corresponds to the Faraday rotation in a cold plasma. The value of $\Delta \phi_{\perp}$ is too small to be easily detectable in simulations with experimentally relevant parameters, i.e. $\omega_{ce} / \omega_* = \left( \omega_{ce} / \omega_* \right)_{exp}$. We thus deliberately set the ratio of $\omega_{ce} / \omega_{*}$ to $\left( \omega_{ce} / \omega_* \right)_{sim} \approx 0.15$ to increase $\Delta \phi_{\perp}$ and make it detectable. In order to make the comparison between $\Delta \phi_{\parallel}$ and $\Delta \phi_{\perp}$ relevant to the experimental setup that we are investigating, we multiply the calculated values of $\Delta \phi_{\perp}$ by $\left( \omega_{ce} / \omega_* \right)_{exp} / \left( \omega_{ce} / \omega_* \right)_{sim}$. The result shown in Fig.~\ref{CM_FR_1D} confirms that in a plasma with relativistic electrons the effect of the magnetic field remains much stronger for the probe propagating parallel to the magnetic field lines.

\section{\sffamily{Faraday rotation for a denser channel\\and for a channel filled with carbon}} \label{App_B}

In the main text, we optimize the radius of a structured target for different laser focusing configurations. In all of the presented simulations, the channel is filled with fully ionized hydrogen whose corresponding electron density is $n_e = 20 n_{cr}$ (see Table \ref{table_PIC}). We have performed two additional 3D PIC simulations to determine what we should expect for a different channel material and a different channel density.

\begin{figure}[htbp!]
\centering 
\includegraphics[scale=0.4]{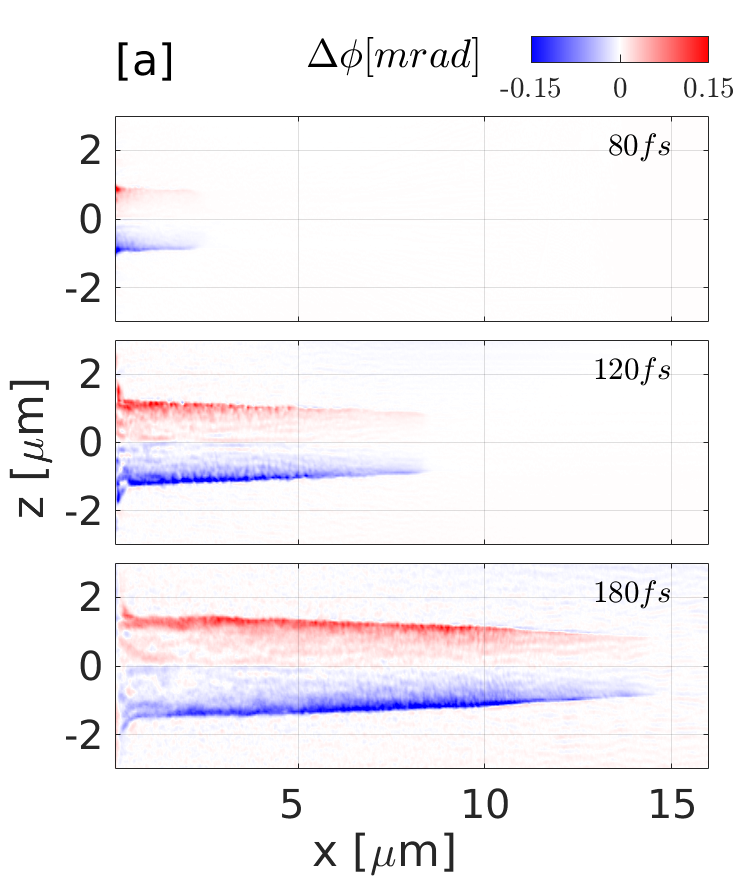} \includegraphics[scale=0.4]{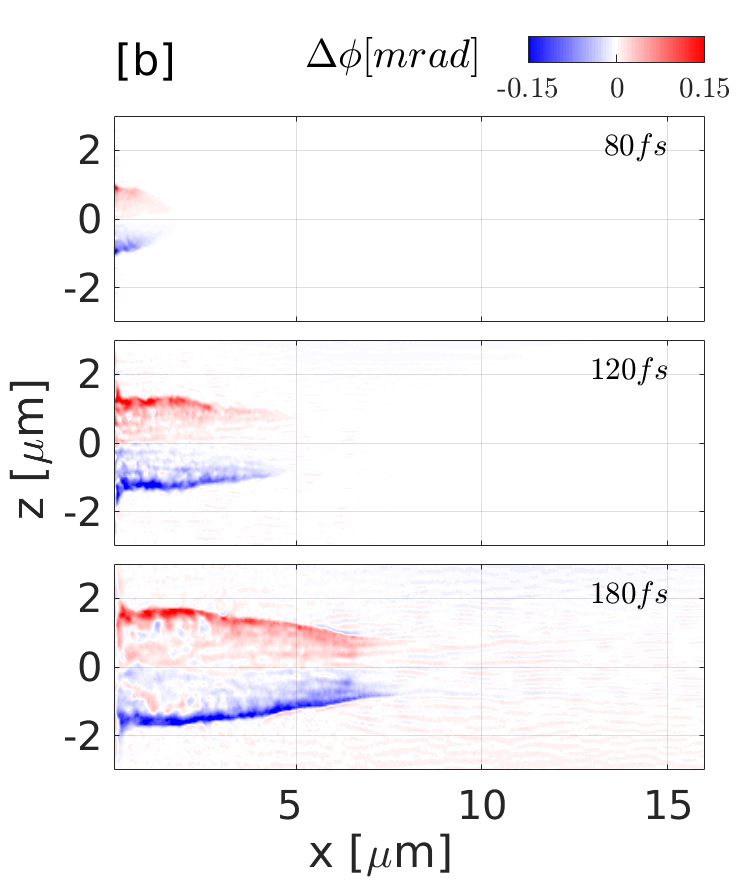} 
\caption{Snapshots of the x-ray beam polarization rotation in the case of the f1 laser focusing configuration detailed in Table \ref{table_PIC}. (a) The rotation in a target whose channel is filled with fully ionized carbon. The electron density in the channel is $n_e = 20 n_{cr}$. (b) The rotation in a target whose channel is filled with fully ionized hydrogen. The electron density in the channel is $n_e = 40 n_{cr}$, which is twice the value listed in Table \ref{table_PIC}.} \label{A2}
\end{figure}

Figure~\ref{A2}a shows three snapshots from a simulation where the channel is filled with fully ionized carbon whose electron density is $n_e = 20 n_{cr}$. The laser pulse and other target parameters are the same as in Table \ref{table_PIC}. The rotation angle amplitude and the size of the detectable area with a strong magnetic field are very similar to those for a channel filled with hydrogen. The corresponding snapshots for a hydrogen-filled channel are not shown because they look almost identical. This indicates that we are not constrained by using hydrogen and another material can be used to fill the channel if that is more practical in terms of target manufacturing techniques. 

Figure~\ref{A2}b shows snapshots from a simulation where the channel is filled with fully ionized hydrogen whose electron density ($n_e = 40 n_{cr}$) is twice the value listed in Table \ref{table_PIC}. All other simulation parameters are the same as in Table \ref{table_PIC}. This denser channel slows down the laser pulse propagation and the magnetic field filament becomes shorter. This is evident from Figure~\ref{A2}b, where the area with the strong rotation is almost two times shorter compared to that in the corresponding snapshots of Fig.~\ref{A2}a. The rotation angle however remains relatively unaffected. This result suggests that the setup can be further optimized by changing the electron density in the channel. A detailed parameter scan is required to determine the optimal electron density in the channel.



\bibliographystyle{ieeetr}
\bibliography{RT.bib}

\begin{thebibliography}{10}

\bibitem{MV_x_rays_CT}
K.~Ruchala, G.~Olivera, E.~Schloesser, and T.~Mackie, ``Megavoltage {CT} on a
  tomotherapy system,'' {\em Physics in Medicine \& Biology}, vol.~44, no.~10,
  p.~2597, 1999.

\bibitem{MV_x_rays_therapy}
W.~Duncan and P.~Quilty, ``The results of a series of 963 patients with
  transitional cell carcinoma of the urinary bladder primarily treated by
  radical megavoltage {X}-ray therapy,'' {\em Radiotherapy and Oncology},
  vol.~7, no.~4, pp.~299--310, 1986.

\bibitem{NRF_3MeV}
E.~Kwan, G.~Rusev, A.~S. Adekola, F.~D\"onau, S.~L. Hammond, C.~R. Howell,
  H.~J. Karwowski, J.~H. Kelley, R.~S. Pedroni, R.~Raut, A.~P. Tonchev, and
  W.~Tornow, ``Discrete deexcitations in $^{235}\mathrm{U}$ below 3 {MeV} from
  nuclear resonance fluorescence,'' {\em Phys. Rev. C}, vol.~83, p.~041601, Apr
  2011.

\bibitem{AI_homeland}
{\em American National Standard Minimum Performance Criteria for Active
  Interrogation Systems Used for Homeland Security}, ch.~ANSI N42.41-2007,
  pp.~1--35.
\newblock 2008.

\bibitem{Ribeye_pair_creation}
X.~Ribeyre, E.~d'Humi\`eres, O.~Jansen, S.~Jequier, V.~T. Tikhonchuk, and
  M.~Lobet, ``Pair creation in collision of $\ensuremath{\gamma}$-ray beams
  produced with high-intensity lasers,'' {\em Phys. Rev. E}, vol.~93,
  p.~013201, Jan 2016.

\bibitem{Chen_MeV_x_ray}
S.~Chen, N.~D. Powers, I.~Ghebregziabher, C.~M. Maharjan, C.~Liu, G.~Golovin,
  S.~Banerjee, J.~Zhang, N.~Cunningham, A.~Moorti, S.~Clarke, S.~Pozzi, and
  D.~P. Umstadter, ``Mev-energy x rays from inverse compton scattering with
  laser-wakefield accelerated electrons,'' {\em Phys. Rev. Lett.}, vol.~110,
  p.~155003, Apr 2013.

\bibitem{MeV_nonlinear_Thomson}
G.~Sarri, D.~J. Corvan, W.~Schumaker, J.~M. Cole, A.~Di~Piazza, H.~Ahmed,
  C.~Harvey, C.~H. Keitel, K.~Krushelnick, S.~P.~D. Mangles, Z.~Najmudin,
  D.~Symes, A.~G.~R. Thomas, M.~Yeung, Z.~Zhao, and M.~Zepf, ``Ultrahigh
  brilliance multi-{MeV} $\ensuremath{\gamma}$-ray beams from nonlinear
  relativistic thomson scattering,'' {\em Phys. Rev. Lett.}, vol.~113,
  p.~224801, Nov 2014.

\bibitem{MeV_Geddes2015}
C.~G. Geddes, S.~Rykovanov, N.~H. Matlis, S.~Steinke, J.-L. Vay, E.~H. Esarey,
  B.~Ludewigt, K.~Nakamura, B.~J. Quiter, C.~B. Schroeder, {\em et~al.},
  ``Compact quasi-monoenergetic photon sources from laser-plasma accelerators
  for nuclear detection and characterization,'' {\em Nuclear Instruments and
  Methods in Physics Research Section B: Beam Interactions with Materials and
  Atoms}, vol.~350, pp.~116--121, 2015.

\bibitem{jll_photon}
L.~Ji, A.~Pukhov, E.~Nerush, I.~Y. Kostyukov, B.~Shen, and K.~Akli, ``Energy
  partition, $\gamma$-ray emission, and radiation reaction in the near-quantum
  electrodynamical regime of laser-plasma interaction,'' {\em Physics of
  Plasmas}, vol.~21, no.~2, p.~023109, 2014.

\bibitem{ELI_NP}
S.~Gales, K.~A. Tanaka, D.~L. Balabanski, F.~Negoita, D.~Stutman, O.~Tesileanu,
  C.~A. Ur, D.~Ursescu, I.~Andrei, S.~Ataman, M.~O. Cernaianu, L.~D?Alessi,
  I.~Dancus, B.~Diaconescu, N.~Djourelov, D.~Filipescu, P.~Ghenuche, D.~G.
  Ghita, C.~Matei, K.~Seto, M.~Zeng, and N.~V. Zamfir, ``The extreme light
  infrastructure?nuclear physics ({ELI-NP}) facility: new horizons in physics
  with 10 {PW} ultra-intense lasers and 20 {MeV} brilliant gamma beams,'' {\em
  Reports on Progress in Physics}, vol.~81, no.~9, p.~094301, 2018.

\bibitem{Stark-PhysRevLett.116.185003}
D.~Stark, T.~Toncian, and A.~Arefiev, ``Enhanced multi-mev photon emission by a
  laser-driven electron beam in a self-generated magnetic field,'' {\em
  Physical Review Letters}, vol.~116, no.~18, p.~185003, 2016.

\bibitem{Oliver_pair}
O.~Jansen, T.~Wang, D.~J. Stark, E.~d?Humi\'eres, T.~Toncian, and A.~V.
  Arefiev, ``Leveraging extreme laser-driven magnetic fields for gamma-ray
  generation and pair production,'' {\em Plasma Physics and Controlled Fusion},
  vol.~60, no.~5, p.~054006, 2018.

\bibitem{LCLS_2010}
P.~Emma, R.~Akre, J.~Arthur, R.~Bionta, C.~Bostedt, J.~Bozek, A.~Brachmann,
  P.~Bucksbaum, R.~Coffee, F.-J. Decker, {\em et~al.}, ``First lasing and
  operation of an {\aa}ngstrom-wavelength free-electron laser,'' {\em nature
  photonics}, vol.~4, no.~9, p.~641, 2010.

\bibitem{XFEL_review}
C.~Pellegrini, ``X-ray free-electron lasers: from dreams to reality,'' {\em
  Physica Scripta}, vol.~2016, no.~T169, p.~014004, 2017.

\bibitem{EUXFEL}
See http://dx.doi.org/10.3204/XFEL.EU/TR-2011-001 for European XFEL parameters.

\bibitem{Japan_XFEL}
T.~Ishikawa, H.~Aoyagi, T.~Asaka, Y.~Asano, N.~Azumi, T.~Bizen, H.~Ego,
  K.~Fukami, T.~Fukui, Y.~Furukawa, {\em et~al.}, ``A compact x-ray
  free-electron laser emitting in the sub-{\aa}ngstr{\"o}m region,'' {\em
  nature photonics}, vol.~6, no.~8, p.~540, 2012.

\bibitem{hibef}
See http://www.hibef.eu for HiBEF project information.

\bibitem{Huang2017}
L.~G. Huang, H.-P. Schlenvoigt, H.~Takabe, and T.~E. Cowan, ``Ionization and
  reflux dependence of magnetic instability generation and probing inside
  laser-irradiated solid thin foils,'' {\em Physics of Plasmas}, vol.~24,
  no.~10, p.~103115, 2017.

\bibitem{law2016_kT_proton}
K.~Law, M.~Bailly-Grandvaux, A.~Morace, S.~Sakata, K.~Matsuo, S.~Kojima,
  S.~Lee, X.~Vaisseau, Y.~Arikawa, A.~Yogo, {\em et~al.}, ``Direct measurement
  of kilo-tesla level magnetic field generated with laser-driven capacitor-coil
  target by proton deflectometry,'' {\em Applied Physics Letters}, vol.~108,
  no.~9, p.~091104, 2016.

\bibitem{santos2015_800T_proton_FR}
J.~Santos, M.~Bailly-Grandvaux, L.~Giuffrida, P.~Forestier-Colleoni,
  S.~Fujioka, Z.~Zhang, P.~Korneev, R.~Bouillaud, S.~Dorard, D.~Batani, {\em
  et~al.}, ``Laser-driven platform for generation and characterization of
  strong quasi-static magnetic fields,'' {\em New Journal of Physics}, vol.~17,
  no.~8, p.~083051, 2015.

\bibitem{Schumaker_PRL_2013}
W.~Schumaker, N.~Nakanii, C.~McGuffey, C.~Zulick, V.~Chyvkov, F.~Dollar,
  H.~Habara, G.~Kalintchenko, A.~Maksimchuk, K.~A. Tanaka, A.~G.~R. Thomas,
  V.~Yanovsky, and K.~Krushelnick, ``Ultrafast electron radiography of magnetic
  fields in high-intensity laser-solid interactions,'' {\em Phys. Rev. Lett.},
  vol.~110, p.~015003, Jan 2013.

\bibitem{stamper1975_seminal_faraday}
J.~Stamper and B.~Ripin, ``Faraday-rotation measurements of megagauss magnetic
  fields in laser-produced plasmas,'' {\em Physical Review Letters}, vol.~34,
  no.~3, p.~138, 1975.

\bibitem{willi1998}
M.~Borghesi, A.~J. Mackinnon, R.~Gaillard, O.~Willi, A.~Pukhov, and
  J.~Meyer-ter Vehn, ``Large quasistatic magnetic fields generated by a
  relativistically intense laser pulse propagating in a preionized plasma,''
  {\em Phys. Rev. Lett.}, vol.~80, pp.~5137--5140, Jun 1998.

\bibitem{kaluzaPRL2010}
M.~C. Kaluza, H.-P. Schlenvoigt, S.~P.~D. Mangles, A.~G.~R. Thomas, A.~E.
  Dangor, H.~Schwoerer, W.~B. Mori, Z.~Najmudin, and K.~M. Krushelnick,
  ``Measurement of magnetic-field structures in a laser-wakefield
  accelerator,'' {\em Phys. Rev. Lett.}, vol.~105, p.~115002, Sep 2010.

\bibitem{walton2013_FR_experiments_abel}
B.~Walton, A.~Dangor, S.~P. Mangles, Z.~Najmudin, K.~Krushelnick, A.~G.~R.
  Thomas, S.~Fritzler, and V.~Malka, ``Measurements of magnetic field
  generation at ionization fronts from laser wakefield acceleration
  experiments,'' {\em New Journal of Physics}, vol.~15, no.~2, p.~025034, 2013.

\bibitem{cecchetti2009_45T_proton}
C.~Cecchetti, M.~Borghesi, J.~Fuchs, G.~Schurtz, S.~Kar, A.~Macchi,
  L.~Romagnani, P.~Wilson, P.~Antici, R.~Jung, {\em et~al.}, ``Magnetic field
  measurements in laser-produced plasmas via proton deflectometry,'' {\em
  Physics of Plasmas}, vol.~16, no.~4, p.~043102, 2009.

\bibitem{Siddons1990_x_ray_Faraday}
D.~P. Siddons, M.~Hart, Y.~Amemiya, and J.~B. Hastings, ``X-ray optical
  activity and the {Faraday} effect in cobalt and its compounds,'' {\em Phys.
  Rev. Lett.}, vol.~64, pp.~1967--1970, Apr 1990.

\bibitem{ji2014_Megatesla}
L.~L. Ji, A.~Pukhov, I.~Y. Kostyukov, B.~F. Shen, and K.~Akli,
  ``Radiation-reaction trapping of electrons in extreme laser fields,'' {\em
  Phys. Rev. Lett.}, vol.~112, p.~145003, Apr 2014.

\bibitem{qiao2017_Megatesla}
B.~Qiao, H.~Chang, Y.~Xie, Z.~Xu, and X.~He, ``Gamma-ray generation from
  laser-driven electron resonant acceleration: In the non-{QED} and the {QED}
  regimes,'' {\em Physics of Plasmas}, vol.~24, no.~12, p.~123101, 2017.

\bibitem{FR_rel2}
R.~V. Shcherbakov, ``Propagation effects in magnetized transrelativistic
  plasmas,'' {\em The Astrophysical Journal}, vol.~688, no.~1, p.~695, 2008.

\bibitem{FR_rel3}
L.~Huang and R.~V. Shcherbakov, ``Faraday conversion and rotation in uniformly
  magnetized relativistic plasmas,'' {\em Monthly Notices of the Royal
  Astronomical Society}, vol.~416, no.~4, pp.~2574--2592, 2011.

\bibitem{FR_rel4}
Y.-P. Li, F.~Yuan, and F.-G. Xie, ``Exploring the accretion model of {M87 and
  3C 84} with the {Faraday} rotation measure observations,'' {\em The
  Astrophysical Journal}, vol.~830, no.~2, p.~78, 2016.

\bibitem{FR_Trubnikov_phd}
B.~{Trubnikov}, {\em {Magnetic Emission of High Temperature Plasma}}.
\newblock PhD thesis, Dissertation, Moscow (US-AEC Tech.~Inf.~Service,
  AEC-tr-4073 [1960]), (1958), 1958.

\bibitem{FR_melrose_1997}
D.~B. Melrose, ``Covariant form of {Trubnikov's} response tensor for a
  relativistic magnetized thermal plasma,'' {\em Journal of Plasma Physics},
  vol.~57, no.~2, p.~479?488, 1997.

\bibitem{MARX2011915}
B.~Marx, I.~Uschmann, S.~H{\"o}fer, R.~L{\"o}ttzsch, O.~Wehrhan,
  E.~F{\"o}rster, M.~Kaluza, T.~St{\"o}hlker, H.~Gies, C.~Detlefs, T.~Roth,
  J.~H{\"a}rtwig, and G.~Paulus, ``Determination of high-purity polarization
  state of {X-rays},'' {\em Optics Communications}, vol.~284, no.~4, pp.~915 --
  918, 2011.

\bibitem{marx2013_xray_polarimetry}
B.~Marx, K.~Schulze, I.~Uschmann, T.~K{\"a}mpfer, R.~L{\"o}tzsch, O.~Wehrhan,
  W.~Wagner, C.~Detlefs, T.~Roth, J.~H{\"a}rtwig, {\em et~al.},
  ``High-precision x-ray polarimetry,'' {\em Physical review letters},
  vol.~110, no.~25, p.~254801, 2013.

\bibitem{Epoch}
T.~Arber, K.~Bennett, C.~Brady, A.~Lawrence-Douglas, M.~Ramsay, N.~Sircombe,
  P.~Gillies, R.~Evans, H.~Schmitz, A.~Bell, {\em et~al.}, ``Contemporary
  particle-in-cell approach to laser-plasma modelling,'' {\em Plasma Physics
  and Controlled Fusion}, vol.~57, no.~11, p.~113001, 2015.

\bibitem{Seedme2}
A.~Chourasia, D.~Nadeau, and M.~Norman, ``Seedme: Data sharing building
  blocks,'' {\em Proceedings of the Practice and Experience in Advanced
  Research Computing 2017 on Sustainability, Success and Impact (PEARC17)},
  vol.~69, p.~1, 2017.

\bibitem{juttner1911maxwellsche}
F.~J{\"u}ttner, ``Das {Maxwellsche} {Gesetz} der geschwindigkeitsverteilung in
  der relativtheorie,'' {\em Annalen der Physik}, vol.~339, no.~5,
  pp.~856--882, 1911.

\bibitem{rezzolla2013relativistic}
L.~Rezzolla and O.~Zanotti, {\em Relativistic hydrodynamics}.
\newblock Oxford University Press, 2013.

\bibitem{zenitani2015loading}
S.~Zenitani, ``Loading relativistic {Maxwell} distributions in particle
  simulations,'' {\em Physics of Plasmas}, vol.~22, no.~4, p.~042116, 2015.

\bibitem{book_FR_CM_cold}
H.-J. Hartfu{\ss} and T.~Geist, {\em Fusion Plasma Diagnostics with mm-waves:
  an introduction}.
\newblock John Wiley \& Sons, 2013.

\end{thebibliography}

\end{document}